# Chapter 6

# Volatiles

Kathleen Mandt[1], Oleksandra Ivanova[2,3,4], Olga Harrington Pinto[5], Nathan X. Roth[1,6], Darryl Z. Seligman[7,8]

[1]NASA Goddard Space Flight Center, Greenbelt, MD, USA
[2]Astronomical Institute of the Slovak Academy of Sciences, Slovakia
[3]Main Astronomical Observatory of the National Academy of Sciences of Ukraine, Ukraine
[4]Taras Shevchenko National University of Kyiv, Astronomical Observatory, Ukraine
[5] Department of Physics, Auburn University, Auburn, AL, USA
[6] Department of Physics, American University, Washington, DC, USA
[7]Department of Physics and Astronomy, Michigan State University, East Lansing, MI, 48824, USA
[8]Department of Astronomy and Carl Sagan Institute, Cornell University, 122 Sciences Drive, Ithaca, NY, 14853, USA

Small bodies are the remnant building blocks from the time when the planets formed and migrated to their current positions. Their volatile composition and relative abundances serve as time capsules for the formation conditions in the protosolar nebula. By constraining the volatile composition of Centaurs, we can fill in important gaps in understanding the history of our solar system. We review here the state of knowledge for volatiles in small bodies, processes that influence volatile composition and activity in small bodies, and future capabilities that can be leveraged to advance our understanding of volatiles in Centaurs.

## 6.1 Introduction

Determining how our solar system formed and evolved is critical for answering several of the fundamental questions identified in the most recent planetary science decadal survey, Origins Worlds and Life (OWL; NASEM, 2022). Furthermore, figuring out why our solar system architecture – or the number, size, and locations of the planets and small bodies – appears to be unique compared to exoplanet and planetary disk systems is a top priority for investigation recognized in the most recent astrophysics decadal survey, Pathways to Discovery in Astronomy and Astrophysics for the 2020s (Astro2020; NASEM, 2020).

Our solar system's architecture is the result of the timing and location of the giant planets' formation and their subsequent migration. We know from exoplanetary systems hosting hot Jupiters that inward migration of Jupiter could have destroyed the terrestrial

planets — or at least removed them from their current day orbits (Fogg & Nelson, 2007a; 2007b; Walsh et al. 2011; Madhusudhan et al. 2014). Furthermore, it is possible that the migration of the giant planets is responsible for delivering volatiles to the inner planets (Raymond & Izidoro, 2017; Kane et al. 2021; see also Chapter 15 Telus et al. *this book*). Therefore, understanding the formation conditions, dynamical history, current distribution, and volatile composition of small bodies in our solar system is critical for determining how our solar system formed and evolved. There are four primary populations of small bodies within the solar system – asteroids, Centaurs, Transneptunian objects (TNOs), and Oort Cloud objects. Comets are a class of small body that originate as both TNOs and Oort Cloud objects. At the present time, the bulk volatile composition of Centaurs as a population is the least understood.

## 6.1.1 Volatiles and solar system formation

Dynamical models for solar system formation have so far provided a range of solutions that are able reproduce solar system architecture constraints, including the current locations of the planets; the distribution of small bodies throughout the solar system; and the small mass of Mars. However, they provide a wide range of potential formation locations and overall migration scenarios for the giant planets.

Early studies of orbital migration suggested that proto-Uranus and proto-Neptune could smoothly migrate 5-10 AU while embedded in the disk (Fernandez & Ip, 1984), allowing the giant planets to form closer together with Saturn, Uranus, and Neptune migrating outward while Jupiter migrated inward (Hahn & Malhotra 1999). In the "Nice" model – initially developed in Nice, France – the giant planets formed within a range of ~5.5–17 AU with a population of small icy planetesimals distributed between ~17 and ~35 AU. Some of these small bodies interacted with Saturn, Uranus, and Neptune and were scattered inward to the inner solar system (Tsiganis et al. 2005) while the planets migrated outward. This outward migration created a dynamical instability that led to the current structure of the Kuiper Belt (Levison et al. 2007). Jupiter's interaction with the small bodies that were scattered inward then caused them to be ejected outward into the Oort Cloud or out of the solar system (Tsiganis et al. 2005; Levison et al. 2007). Neptune is also thought to have scattered bodies outward. Alternatively, the "Grand Tack" model proposes that Jupiter formed at 3.5 AU and migrated inward to 1.5 or 2 AU (Walsh et al. 2011; Brasser et al. 2016a,b). Saturn formed at a slower rate, then migrated inward at a higher velocity (e.g. Masset and Papaloizou 2003) allowing it to catch up to and capture Jupiter in a 3:1 resonance (Masset and Snellgrove 2001; Morbidelli and Crida 2007; Pierens and Nelson 2008). After this, both giant planets migrated outward to their current locations. In this scenario Jupiter crossed the asteroid belt twice leading to its current mass and distribution (Walsh et al. 2011). More recent studies suggest that a fifth giant planet, potentially another ice giant, formed as well but was ejected from the solar system (e.g. Deienno et al. 2017; Nesvorny, 2011). Current observational constraints are not sufficient to determine which dynamical model scenario provides the best solution.

Models for how the giant planets formed have also attempted to constrain their formation locations, with little success. Originally, each planet's core was assumed to have grown hierarchically from small grains to a planet many times the mass of Earth. However, the current locations of Uranus and Neptune did not have sufficient solid material available at that distance. Therefore, they either must have formed closer to the Sun, as in the Nice model, or formed through a different method. The streaming instability offers a promising alternative, where collective aerodynamic interactions in the protosolar nebula (PSN) produced high density filaments and streams that allowed pebbles to form (Youdin and Goodman, 2005; Johansen & Youdin, 2007; Johansen et al. 2015; Simon et al., 2016) and clouds of pebbles to contract into planetesimals (Nesvorny et al. 2019). This mechanism permits formation at greater distances given sufficient dust to gas ratios. Pebble accretion also works at greater distances, in which case pebbles are slowed by gas drag in the PSN and rapidly accumulate onto existing planetesimals (Lambrechts & Johansen, 2014). For a detailed review see Chapter 2 of this book (Johansen et al, *this book*). These models, therefore, are not able to limit the formation locations of the giant planets to a specific region of the PSN.

The remaining approach for constraining solar system formation and evolution is to compare models for the volatile composition of the solid building blocks of the planets as a function of time and distance in the PSN with the composition of small body populations and of the giant planet atmospheres. Combining this with dynamical models can help to identify the formation locations and times for the giant planets and for the various small body populations remaining today. Recent modeling evaluated the influence of ice lines of specific species in the PSN and the type of water ice to determine the composition of Jupiter's atmosphere based on when and where it formed (Mousis et al. 2019; Aguichine et al. 2022; Schneeberger et al. 2023). Although these models provide new information for understanding the formation processes of solid building blocks, the observational constraints remain limited. Future observations of the bulk elemental composition of the giant planet atmospheres and small body populations, as well as isotopic ratios of several elements would help to constrain formation locations.

## 6.1.2 Role of Centaurs in understanding solar system formation

We know that small icy bodies, including Centaurs, formed in an extensive region of the PSN with temperatures that could have ranged between 200 K at about 5 AU to approximately 30 K near 40 AU. The models for ice formation in the PSN show that the composition of volatile ices formed within this wide temperature range can vary significantly (Mousis et al. 2019; Aguichine et al. 2022; Schneeberger et al. 2023). Variability in the volatile composition of small body populations could indicate different regions where different populations formed.

Centaurs are proposed to originate as TNOs and to represent a steady-state population that transfers material from the Trans-Neptunian region to the inner solar system eventually as Jupiter Family Comets (JFCs) (Emel'Yanenko 2005; Brasser et al. 2012; Sarid et al. 2019; Di Sisto et al. *this book*). For a time, they reside within the orbits of the giant planets and are subject to repeated dynamical interactions with the giant planets. If this is the case, then their formation location was the same as comets and the initial volatile inventory should be similar. Any differences would be due to volatile evolution over time. If this is not the case, then their volatile composition would indicate their formation region. Constraining the dynamical history of Centaurs compared to TNOs and JFCs requires a comparison of the volatile inventories of Centaurs with what is known of the volatile inventories of JFCs.

In either case, evolution of volatiles needs to be considered. The primary type of evolution would be preferential removal of volatiles that sublimate at lower temperatures than water when the surface and interior are heated. Although studies have been done on volatile evolution in comets (e.g. Guilbert-Lepoutre et al. 2015) and Centaurs (e.g. Fernandez et al. 2018; Kokotanekova et al. *this book*), more work in the form of observations with new state of the art facilities on Earth and in space, as well as laboratory and theoretical studies is required to fully understand how volatile depletion works.

A second question about Centaurs and their volatile inventories is what impact they may have had on the type of volatiles delivered to the inner planets (see Telus et al. *this book*). The inner planets' current atmospheres formed in part from internal degassing, and in part due to the impacts of water-rich small bodies scattered into the inner solar system (Marov and Ipatov, 2001; Schönbächler et al., 2010). The relative contribution of volatiles from icy bodies and asteroids may be constrained by their atmospheres' noble gas content and isotopic composition (e.g. Owen and Bar-Nun, 2001) given sufficient information on small body composition.

## 6.2 Volatiles in small body populations

The bulk volatile composition of different small body populations is constrained by observations of volatiles that are on the surface and volatiles released into a coma through heating and through other processes. Understanding what the observations mean for the bulk composition requires understanding volatile phases and processes that change the phases of volatiles. We outline in this Section the volatile phases relevant to small body populations and describe observations of small body populations including Centaurs.

### 6.2.1 Volatile phases

The ice in small bodies can be present in different phases, or forms, depending on the temperature of formation, the thermal processing experienced over the history of the small body, and any coma activity that leads to redeposition of volatiles. Ice phases relevant to Centaurs include amorphous, condensed, crystalline, and clathrates.

**Amorphous Ice**

Amorphous ice is formed when water molecules freeze rapidly under low-temperature conditions (< 136 K), preventing the formation of a well-defined crystalline structure found in water ice on Earth (Mastrapa et al., 2013). The low temperature conditions where cometary nuclei formed would allow for condensation as amorphous ice that could still be preserved (Pataschnik et al. 1974). However, amorphous ice is unstable thermodynamically and can undergo an exothermic transformation into a crystalline state when exposed to warmer temperatures or additional energy (Bar-Nun et al., 1985). The temperature of the ice influences the rate at which this conversion occurs.

Amorphous ice is ~2 times more dense than crystalline ice and will crystallize when heated to a temperature of 150 K. Crystallization occurs exothermically and rapidly. The temperature rises to 190 K, releasing 0.02 eV per molecule. This energy is insufficient to cause evaporation of the water ice, as shown in Figure 6.1. However, crystallization models depend on poorly constrained parameters leading to uncertainties regarding processes related to amorphous ice in cometary nuclei (Kouchi & Sirono 2001). It is possible that the transition in natural ice might not be exothermic due to the sublimation of trapped volatiles, but because the trapped volatiles are in buried ice, they cannot escape directly to space and the phase transition energy would remain trapped.

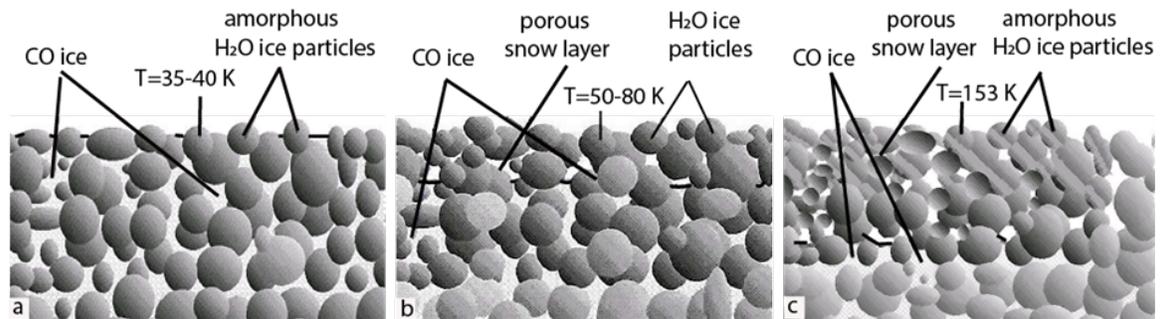

Figure 6.1: Illustration of the process of crystallization of amorphous ice. (a) The initial state of the object's surface with amorphous water ice particles embedded in a CO ice substrate; (b) A surface porous layer forms from microcrystals of amorphous ice at the onset of surface heating; (c) The crystallization front of amorphous ice moves deeper into the nucleus (based on model presented in Shulman 2007).

Crystallization of amorphous ice can occur at lower temperatures over longer timescales. However, at temperatures below 77 K, the crystallization time exceeds the age of the solar system (Notesco & Bar-Nun, 1996; Notesco et al., 2003). The potential for phase transition in a specific body can be estimated by comparing the crystallization time to the orbital period where amorphous ice is expected to have crystallized on objects with crystallization time shorter than the orbital period (Notesco & Bar-Nun, 1996; Notesco et al., 2003).

**Condensed ice**

Volatiles condense to form ice at temperatures and pressures specific to each molecule. The left panel of Fig. 6.2 illustrates the condensation curves for volatiles observed in comets at pressures and temperatures relevant to the PSN. The heavy noble gases, $CH_4$, CO, and $N_2$ condense at the coldest temperatures while $NH_3$, $CO_2$, and $H_2S$ condense at warmer temperatures. Water condenses at temperatures above the range in this figure.

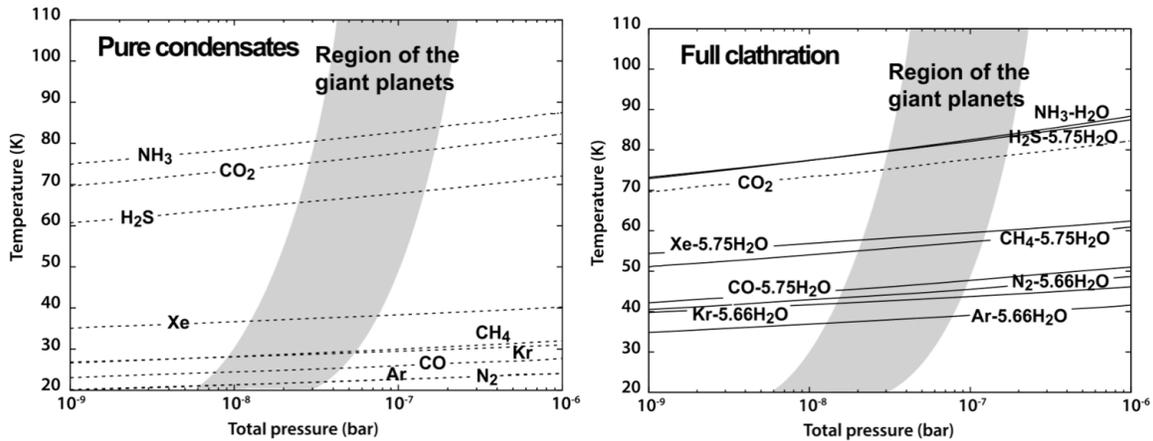

Figure 6.2: Condensation curves serve as a tool for prediction building block composition as a function of conditions in the PSN. Here they are compared to the predicted cooling curve of the PSN where the giant planets may have formed. (left) Equilibrium curves for pure condensates (dashed lines) where species are in the gas phase above the curves. (right) Scenario for clathrate formation including a combination of condensate, hydrate ($NH_3$–$H_2O$), and clathrate formation (X–$5.75H_2O$ or X–$5.67H_2O$; solid lines), along with crystallization of pure $CO_2$ condensate (dotted line). (Mousis et al. 2020) reprinted with permission © Springer.

**Crystalline Ice and Clathrate Hydrates**

Crystalline ice, including clathrate hydrates, is known to form under a range of thermodynamic conditions in small bodies (Kargel & Lunine, 1998). Unlike amorphous ice, which only forms at very low temperatures, crystalline ice forms at temperatures >110 K and has a regular lattice-like structure. It is worth noting that crystalline ice can also be converted to amorphous ice via exposure to high energy radiation, or cosmic rays. This high energy radiation or cosmic rays disrupts the crystalline structure and creates voids and other amorphous properties (Hansen & McCord 2004). Therefore, it is possible that amorphous ice is created and retained in TNOs which are not at sufficient ambient temperatures for the amorphous ice to re-crystallize.

Clathrate hydrates are a unique form of crystalline ice and are compounds where water molecules form a lattice structure with small cages that trap guest molecules, such as methane or carbon dioxide. Two common clathrate structures are structures I and II, characterized by different water cells (Sloan & Koh, 2008). Structure I has pentagonal and tetrakaidekahedral cells, while structure II has pentagonal and hexakaidecahedral cells. The structure selection relies on the dimensions and configuration of the guest molecule. For example, $CH_4$ and $CO_2$ are characteristic of structure I, while $N_2$ favors structure II (Sloan & Koh, 2008). The right panel of Fig. 6.2 illustrates one potential scenario for the formation of crystalline ice and clathrates in small bodies (Mousis et al. 2020).

Early on it was proposed that $CH_4$, $NH_3$, and $CO_2$ in cometary ices exist as clathrate hydrates rather than macroscopic inclusions (Delsemme and Swings (1952) and thermodynamic models suggest that $CH_4$ should always be trapped in clathrates beyond the water snowline (Mousis et al., 2015; Luspay-Kuti et al. 2016). Furthermore, the condensation process would involve the construction of hydrates instead of direct condensation from the gas phase, which could explain the presence of $CH_4$ and $NH_3$ in cometary nuclei because their high vapor pressures would cause immediate evaporation if not trapped in clathrates.

Additionally, thermodynamic calculations suggest that $N_2$ would have condensed as pure ice at lower temperatures than required for clathration, while $CH_4$ and CO would have remained stable (Mousis et al., 2015; Luspay-Kuti et al. 2016). Any radiogenic heating would lead to the devolatilization of $N_2$ while preserving $CH_4$ and CO in cometary nuclei.

**Observations of volatile phases**

Observations regarding the physical state of ice in small bodies is limited. Because amorphous ice is expected to form at the temperatures where TNOs orbit and crystalline ice can be converted back to amorphous ice here over ~10 Myr timescales, it is expected to exist on the surfaces of TNOs. However, the surfaces of some TNOs exhibit reflectance spectra typical of crystalline ice. For example, the crystalline ice on the surface of (50000) Quaoar (Jewitt & Luu 2004) suggests that it was created in the recent past.

Although distinguishing amorphous ice from crystalline ice on the surfaces of active small bodies at large heliocentric distances remains challenging, the presence of clathrates can be inferred. Methane has been observed in several cometary comae while crystalline water ice was detected on the surface of comet 9P/Tempel 1 (Sunshine et al., 2006), both supporting the hypothesis of a methane-rich clathrate layer. Further support has been provided by Rosetta observations of comet 67P/Churyumov-Gerasimenko (67P/C-G) (Mousis et al. 2015; Luspay-Kuti et al. 2016).

While clathrates in comets are actively studied, indirect observational evidence for amorphous ice in Centaurs is found with no evidence for water ice clathrates (Klinger et al., 1996; Devlin 2001). Furthermore, the perihelion distribution and formation conditions of Centaurs suggest that amorphous forms of ice are more likely than clathrates (Jewitt, 2009).

Volatile composition is an important tool for evaluating the form of ice that is present in a small body. For example, the presence of highly volatile ices, such as $N_2$, Ar, CO, and $O_2$ in these objects seems to suggest that they formed at temperatures within the sublimation range of these species, 22–25 K (Klinger et al., 2012; BarNun et al., 1987; Meech & Svoren, 2004). However, it is also possible that these species were trapped in amorphous or crystalline ice. Models of the relative composition of these species in specific ices formed in conditions within the PSN serve as useful tools for evaluating the form of ice in small bodies (Mousis et al. 2019; Aguichine et al. 2022; Schneeberger et al. 2023).

## 6.2.2 How volatiles are released

Volatiles are released by several different mechanisms. Work is ongoing to understand what mechanisms may drive activity in Centaurs compared to comets. We review here three possible mechanisms for volatile release. Depending on the mechanism there may be nongravitational accelerations that influence the dynamics of comets and Centaurs.

**Sublimation**

Sublimation is the primary driver of activity and coma formation in most comets within the solar system. Solar radiation received at the surface of a small body provides energy for a phase transition from ice to gas that produces an outflow of gas. After release, the radiation energy is converted into kinetic energy as the now liberated gas molecules are heated to a thermal speed that is approximately the local sound speed. Dust particles that are trapped within the bulk ice matrix or residing on the surface of the comet are also liberated with the outgassing volatiles, producing a dusty tail.

Sublimation lines, also called snow lines, determine which volatiles are sublimated and are similar to the freeze out lines in protoplanetary disks of various volatiles. Three major snowlines relevant to comet activity are those for $H_2O$, $CO_2$ and $CO$, which are currently located approximately at the orbits of Jupiter, Saturn, and Neptune, respectively. Interior to 4 AU, within the vicinity of the water snow line (or sublimation front), activity is driven mostly by direct sublimation of $H_2O$, and $H_2O$ production rates, $Q$, increase with decreasing distance (A'Hearn et al. 1995). At further distances, past the nominal $H_2O$ snowline, activity has been attributed to direct sublimation of more tenuous volatile species such as $CO_2$ and $CO$.

**Crystallization of Amorphous Ice**

Amorphous ice is highly porous and contains trapped molecules that are released upon crystallization (Bar-Nun et al., 1985), which involves rearranging water molecules into an ordered, crystalline lattice (Mastrapa et al., 2013). The initial step is nucleation *i*, where a small cluster of water molecules organizes itself in a repeating pattern, initiating the growth of a crystal (Kouchi et al., 2002). The presence of impurities, or specific conditions such as mineral surfaces or radiation exposure, can influence nucleation. Surface conditions also affect the conversion process. Solar radiation, for example, can induce structural changes in amorphous ice, facilitating its conversion into the crystalline form. Heliocentric distance, surface roughness, topography, and volatile compounds can further influence crystallization (Gibb et al., 2000; Protopapa et al., 2018).

Crystallization models have been studied extensively for cometary activity. Because this process involves a release of energy, crystallization is suggested to provide additional energy sources that can explain mass loss from comets at low temperatures (Enzian et al., 1997; Prialnik., 1997). Studies of crystallization in relation to cometary activity are also relevant to Centaurs. All active Centaurs have small enough perihelia for the crystallization of amorphous water ice to be a contributing factor to any observed activity, indirectly suggesting the presence of amorphous ice in Centaurs (Notesco & Bar-Nun, 1996). However, translating laboratory experiment results to cometary and Centaur timescales poses challenges due to the vast difference in timescales.

**Clathrate destabilization**

Clathrate hydrates consist of water cages that enclose guest molecules. The characteristics of the guest molecule determine the clathrate structure and impact the stability and subsequent destabilization processes (Sloan & Koh, 2008). Explosive outgassing of clathrates as they become destabilized has been proposed as an explanation for the presence of chaotic terrains found on many icy bodies of the outer solar system.

The stability of clathrates is affected by many factors, including temperature, pressure, composition, and impurities (Mousis et al., 2015). Elevating temperatures can destabilize clathrates, resulting in the liberation of trapped gases. Temperature increases may be caused by solar radiation, internal heat sources, or impact events (Mousis et al., 2015). Reductions in pressure caused by sublimation, outgassing, or shock events can disrupt the clathrate lattice, destabilizing the clathrate and facilitating the release of trapped gases. Irradiation, thermal cycling, and mechanical disruption can also destabilize clathrates in small bodies (Davidsson et al., 2016).

Destabilization of clathrates has long-term implications for the evolution of small bodies. This is because the release of volatiles modifies surface properties and can result in the formation of secondary compounds (Kargel & Lunine, 1998).

### 6.2.3 Activity drivers for Centaurs

The primary driver of activity in Centaurs is thought to be sublimation caused by the Centaur crossing sublimation fronts of volatiles as it approaches perihelion. As noted above, the $H_2O$ sublimation front is located at ~4AU, depending on characteristics of the particular Centaur such as albedo. The sublimation fronts of CO and $CO_2$ are at greater distances than this allowing them to begin sublimating at greater distances from the Sun. The Centaur 29P Schwassmann-Wachmann 1 (hereafter 29P/S-W1) is a prime example of activity initiated by species that are more volatile than water. It has been observed to be active in a state of constant outgassing since its initial discovery. Its semi-major axis of 5.986 AU is interior to the CO ice line and it has a dust coma with CO-dominated outgassing in its recent history (Senay & Jewitt 1994; Crovisier et al. 1995; Gunnarsson et al. 2008; Paganini et al. 2013). Recent observations with the James Webb Space Telescope (JWST) have also detected $CO_2$, finding distinct jets with heterogenous composition (Faggi et al. 2024). Similarly, the Centaurs (60558) 174P/Echeclus and (2060) Chiron display similar (hereafter Echeclus and Chiron), but lower levels of production of CO (Wierzchos et al. 2017, Womack & Stern 1999).

In addition to sublimation, the crystallization of amorphous ice may drive activity in some Centaurs. The subsurface temperatures of Centaurs are consistent with crystallization fronts of amorphous water ice at approximately 10–12 AU (Guilbert-Lepoutre 2012). There is indirect evidence of this process occurring in some active Centaurs, as activity is seen at distances that are too great to correspond to sublimation of $H_2O$. Active Centaurs – defined as having non point source like extended tails – are a small fraction of all of the known Centaurs and have lower mean perihelia distance (5.9AU) than inactive centaurs (8.7AU; Jewitt 2009). This is too far from the Sun for sublimation of $H_2O$ ice to drive activity, while sublimation of CO ice would be *too* efficient to explain the mass loss seen.

This suggests that the crystallization of amorphous water ice was responsible for the activity (Jewitt 2009).

Further evidence for amorphous conversion driving activity on Centaurs is found in spectroscopic measurements of Centaur (10199) Chariklo at different dates indicating both the presence and absence of $H_2O$ surface ice. This behavior was consistent with crystalline ice in the center of the Centaur with amorphous ice on the surface, the crystallization of which could drive the sporadic activity (Guilbert-Lepoutre 2011). Moreover, detailed three-dimensional thermal evolutionary models found that outgassing due to the crystallization of amorphous ice cannot be sustained for greater than ~$10^4$ years, implying that the active Centaurs only recently experienced changes in orbits that triggered activity (Guilbert-Lepoutre 2012).

Impacts of smaller objects with Centaurs could also cause activity by heating volatiles in the region where the impact took place (Wierzchos et al. 2017; Womack et al. 2017). The volatiles released would include those from the impactor and from the Centaur. In this case, the relative abundances of volatiles released from the Centaur would depend on their volatility, for example more CO and $N_2$ would be released than water because they sublimate at much lower temperatures (see discussion in Mandt et al. 2022).

Additionally, the release of icy dust from the nucleus can contribute volatiles into the coma as an extended source. This has been suggested as a source of additional activity for comets C/1995 O1 Hale–Bopp (Cudnik 2005) and 17P/Holmes (de Almeida et al. 2016). Studies of the target comet for the Rosetta mission, 67P/C-G found through observations in the coma that distributed hydrogen halides, HF, HCl, and HBr – species that are known to freeze out on icy grains in molecular clouds – likely originated from icy grains in the dust coma (de Keyser et al. 2017). Some of the CO observed in the coma of Centaur 29P/S-W1 is proposed to originate from icy dust grains in the coma (Gunnarsson, 2003), although observations exist of entirely unrelated dust and CO activity in 29P/S-W1 (Wierzchos et al. 2020). Some water is observed in the coma of 29P/S-W1, with an observed relative production of $Q$CO/$Q$$H_2O$ of ~4.64 at a heliocentric distance of 6.18AU (Ootsubo et al 2012), despite being far outside of the $H_2O$ sublimation front. This water is proposed to originate from icy dust grains (Bockelee-Morvan et al. 2022).

### 6.2.4 Observing volatiles and connecting to bulk composition

As described in Section 6.1, we need to understand the bulk composition of small bodies to connect them to the processes involved in the formation and early evolution of the solar system. We review the methods currently used to observe volatiles in comets and Centaurs, including observations of surface volatile composition and observations of volatiles in the coma. We then discuss how volatiles observed on the surface and in the coma provide detections of specific species that are present, but require further analysis to determine the bulk composition of the body. The task of determining the comprehensive composition is particularly challenging for both comets and Centaurs, given the constraints of limited observational data, underscoring the need for ongoing research in this field.

**Surface Volatiles**

A detailed overview of the surface composition of Centaurs was provided in Chapter 5 (Peixinho et al. *this book*). Volatiles located on the surface can provide some insight into the bulk volatile composition of Centaurs. These volatiles are studied through observations of their surface spectra, surface colors, and through polarimetry.

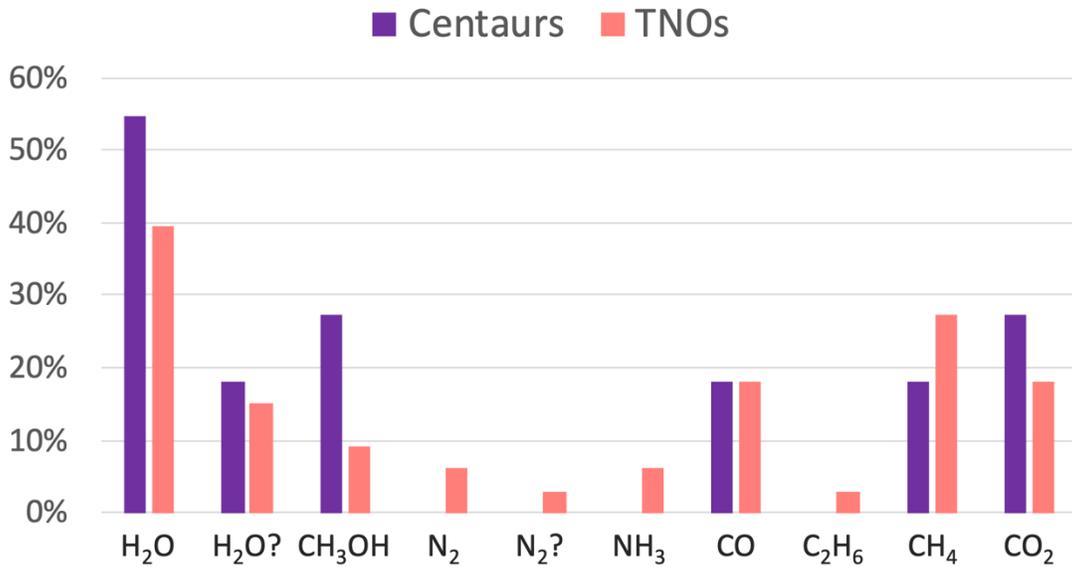

Figure 6.3: Percentage of Centaurs and TNOs on which specific volatile species have been detected (or are suspected to have been detected when "?" is present) in spectral observations of the surface ices. (data combined from de Bergh et al., 2013 and Dalle Ore et al. 2015).

Spectroscopic observations can detect specific ice species, but this can be challenging for distant and faint objects like Centaurs and TNOs (Delsanti et al., 2006; Barucci et al. 2008; Fornasier et al., 2009; Harrington Pinto et al., 2022). Laboratory experiments play a crucial role in interpreting spectra. Detailed studies of crystalline water ice, pure $CH_4$, $CH_4$ diluted in $N_2$, $N_2$ (alpha and beta phases), and CO have been done (Schmitt et al. 2012) as well as $CO_2$ and $CH_3OH$ (Fulchignoni et al. 2008; de Bergh et al., 2013) and irradiation effects on ice and their role in creating organic refractory residues. Fig. 6.3 illustrates the species that have been detected in spectral observations of Centaurs and TNOs, including $CH_4$, $N_2$, and CO surface ice with percentages representing the percent of the objects that have been observed for which each species has been detected so far (de Bergh et al., 2013; Dalle Ore et al. 2015).

In addition to direct detection of surface volatiles, the surface colors of Centaurs may provide insights into their volatile composition. Surface colors are categorized based on how reddened they are, with an index such as B–R colors or spectral slopes. Fig. 6.4 illustrates the B–R color distributions of several small body populations within the solar system, including Centaurs. The majority of small bodies have slightly reddened colors but the cold classical KBOs have an excess of what is commonly referred to as ultra-red material. This ultra-red material presumably results from processing. The colors of Centaurs appear to be bimodal (Jewitt 2009), with a mix of mildly reddened Centaurs and ultra-red Centaurs.

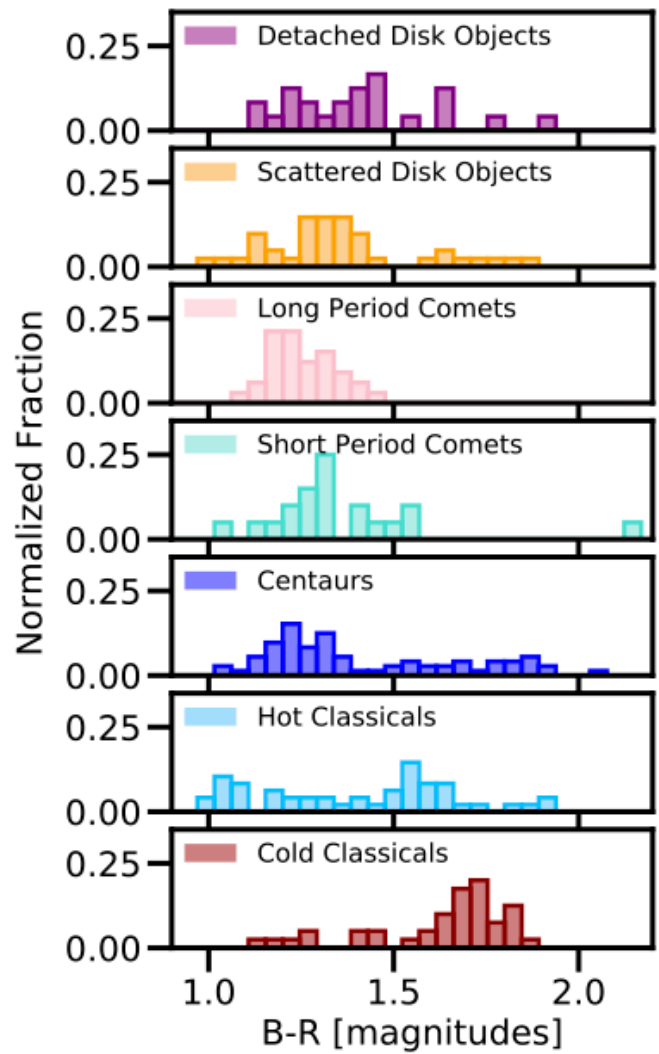

Figure 6.4: Color distributions of various populations of small bodies. Shown on the x-axis for each panel are the B-R measured colors of the surfaces, or the extended comae for active objects. Each panel depicts the color distributions measured in different populations of distant and cometary objects, with Centaurs shown in dark blue (data from Hainaut et al. 2012).

Fig. 6.5 illustrates the inclination distributions of red and blue Centaurs, including both active and inactive Centaurs. It would appear feasible that the redder Centaurs could come from reservoirs with more reddened material, specifically the cold classical KBOs, which is the only solar system small body reservoir that also has ultra-red material (see Fig. 6.4). However, if the redder Centaurs originated as cold classical KBOs, this should be evident in their inclination distributions. Because there is no obvious dependence on the inclination distributions for either the red or the blue centaurs, and because the cold classical KBOs are suggested by long term dynamical simulations to be dynamically very stable (Nesvorny 2017), the colors of the Centaurs are not likely to correlate with the colors of their formation reservoirs.

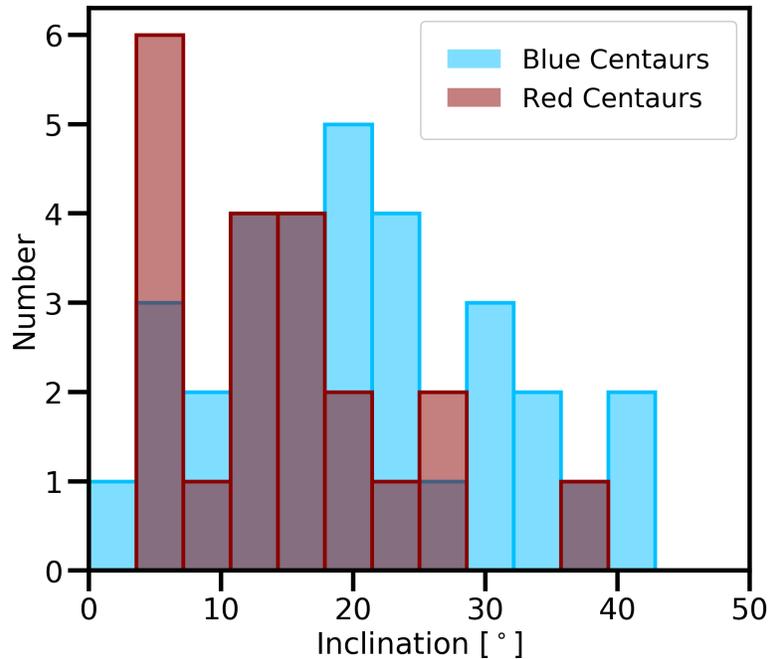

Figure 6.5: Inclination distribution of Centaurs sorted by their colors. The "red" centaurs have B–R > 1.5 and the "blue" have B–R < 1.5. These include active and inactive Centaurs. The data is drawn from Hainaut et al. 2012 and the orbital elements (inclination in this case) are drawn from the Minor Planet Center database.

As Fig. 6.6 illustrates, inactive Centaurs have a bimodal distribution including some ultra-red material and some bluer material, while active Centaurs are blue with the exception of 523676 (2013 UL10) (Mazzotta Epifani et al. 2018). The possible interpretation is that activity removes reddened material from the surface of active centaurs (Jewitt 2015), although the details are not clear given that there exists observations of one reddened active centaur to date. Further studies to explain the color distribution and determine any connection to volatile composition are clearly needed.

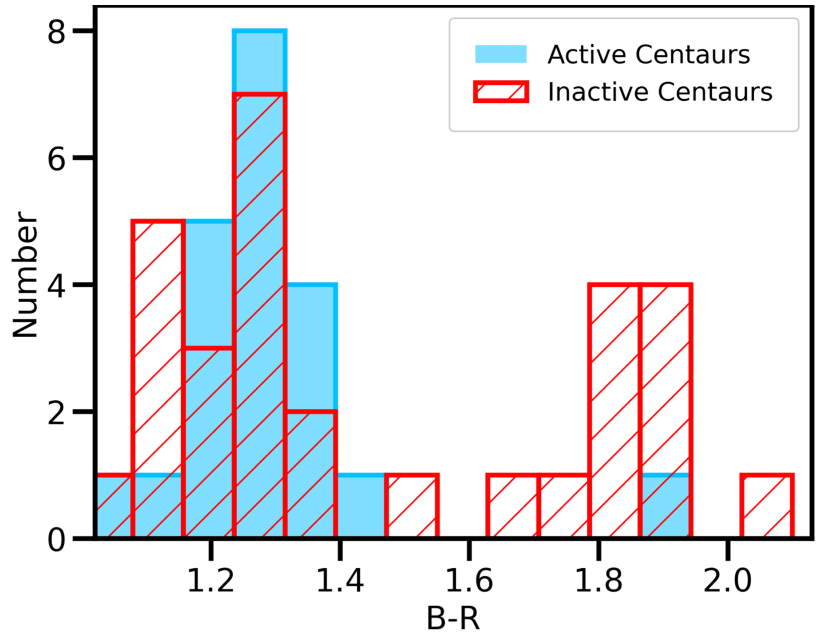

Figure 6.6: Photometric colors of active and inactive centaurs. Data is taken from Jewitt 2015. The blue histograms show measured B–R colors of active centaurs, while the red hatched histogram show the B–R colors of inactive Centaurs. The active red centaur is 523676 (2013 UL10) from Mazzotta Epifani 2018.

Polarimetric observations provide useful insights into the surface properties, composition, and dust dynamics of Centaurs. When an active Centaurs is producing a coma, the sublimation of volatile releases dust and gas, and the coma can contain ice particles. This influences the scattering properties of light, leading to different polarization signatures for the coma compared to the surfaces of Centaurs. Additionally, activity may change the surface polarization signatures of active Centaurs when they are observed during inactive time periods.

Observations have been made of the surfaces of most Centaurs during inactive periods, with the exception of 29P/S-W1 which is found to be active in all observations. Deep negative polarization has been observed in smaller-sized Centaurs, and one possible explanation for this is the presence of an optically heterogeneous surface with dark and bright scatter regions, such as a dark surface covered by a very thin layer of submicron water crystals. The observed variation in linear polarization (P, %) with phase angle between active and inactive Centaurs (see Fig. 6.7 Kiselev and Chernova, 1979; Kochergin et al. 2021) may be related to differences in the abundance of surface ices and dust properties on their surfaces. Results of polarimetric observations in specific active and non-active Centaurs are provided in Section 6.2.5.

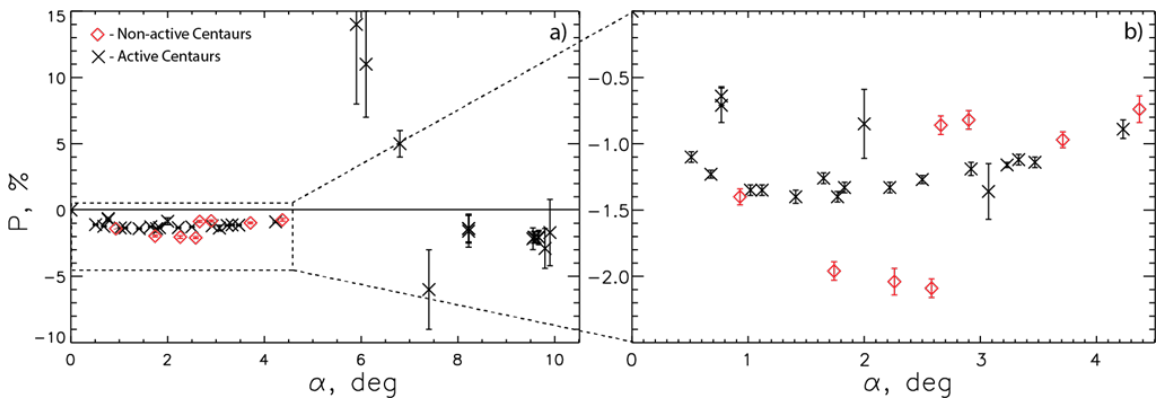

Fig. 6.7. The phase-angle dependency of linear polarization degree of Centaurs provides useful insights into the surface properties, composition, and dust dynamics. This is an illustration of polarization measurements of Centaur surfaces during non-active time periods measured in the R band (a) Polarization vs. phase angle, excluding 29P/S-W1. (b) Close-in view.

**Coma Morphology**

Active Centaurs are considered transitionary between asteroids and comets, showing comet-like activity and an asteroidal nature. Currently activity in ~20 Centaurs has been detected in the form of a dust or gas coma. The composition of their comae can be analyzed through spectroscopic observations, providing insights into the nucleus composition and its evolution.

The morphology of the coma in active Centaurs provides insights into the underlying physical processes and dynamical behavior. We describe here the morphology and evolution of comae in active Centaurs as they relate to volatile composition. Evolutionary processes and Centaur activity are described in greater detail in Chapters 7 & 8 (Kokotanekova et al., *this book*; Bauer et al. *this book*).

Morphological structures (tails, jets, fans, shells, arcs, etc.) of active Centaurs can be formed by dust, neutral gas, and ions, and indicate processes that drive activity (Ivanova et al., 2016; Picazzio et al., 2019; Korsun et al., 2008, Miles et al., 2016a, b; Rousselot et al., 2016; 2021; Kulyk et al., 2016). Activity can be short- or long-term, periodic or spontaneous and features can be symmetric or asymmetric.

The coma morphology of 29P/S-W1 is highly variable. The total coma magnitude brightens for a few days after an outburst, likely due to subsequent sublimation and fragmentation of particles (Trigo-Rodríguez et al., 2008; 2010). Outburst comae generally fade over weeks, but sometimes a new condensed coma appears after another outburst (Miles 2016b). Asymmetric, fan-shaped comae, sometimes described as radial jets, suggest dust grain acceleration by volatile ice sublimation and imply continuous feeding from an active source on the nucleus. Four jet-like structures observed over long time periods in 29P/S-W1 (Shubina et al., 2023) suggest that the coma is formed by active regions located within a narrow belt near the equator. Additionally, $CO^+$ emission and continuum revealed differing distributions of dust and $CO^+$ ions in the coma, depending on the comet's activity level (see Fig. 6.8 for $CO^+$) (Ivanova et al., 2019). The $CO^+$ ions were more concentrated towards the nucleus than the dust continuum. Notably $CO^+$ production appears to be correlated with solar wind intensity suggesting that the solar wind could be the dominant source for ion production at this distance from the Sun compared to closer in comets where ion production is dominated by photoionization (Ivanova et al., 2019). Finally, recent JWST observations found separate CO and $CO_2$ jets that indicate possible differences in nucleus composition (Faggi et al. 2024).

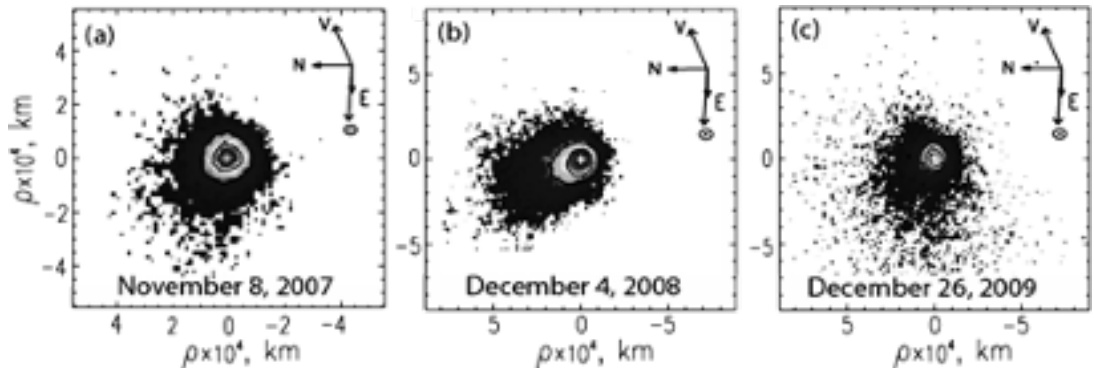

Figure 6.8: Observations of how the morphology of 29P/S-W1 has changed over time based on $CO^+$ in the coma: (a) November 8, 2007, (b) December 4, 2008 and (c) December 26, 2009. Produced using data from Ivanova et al. (2019).

Some Centaurs demonstrate unpredictable outbursts that are not correlated with their perihelion passage. Echeclus is an interesting case. Its primary outburst, which lasted several months, was detected during its pre-perihelion phase at a distance of 13 AU. An apparent source seemed to move away from its nucleus, initially thought to be caused by impacts and fragmentation, but observations did not support this. Three smaller subsequent outbursts were also observed. Notably, observations revealed an unusual blue dust coma, possibly indicating a carbon-rich dust composition. A faint emission line of CO was detected at a distance of 6 AU from the Sun (Wierzchos et al., 2017).

**Coma volatile composition**

The coma of an active small body provides direct access to volatiles in the nucleus, whether in ice form in the nucleus or condensed onto dust that is also released into the coma. The volatile ices are thought to primarily consist of $H_2O$ ice with additions of macroscopic inclusions of ices such as CO, $CO_2$, $CH_4$, and other ices such as $CH_3OH$, $NH_3$, HCN, etc. (Barucci et al., 2008; Wierzchos and Womack, 2020; Harrington Pinto et al., 2022). However, connecting observations in the coma to the bulk composition of a comet or Centaur is challenging because the composition of a coma varies as a function of the orbit due to changes in the temperature of the nucleus. The most comprehensive evaluation so far has been done for the comet 67P/C-G using observations from Rosetta (Rubin et al. 2019).

Neutral species observed in a coma fall into three categories: (1) "parent" or "primary" species that sublimed directly from the nucleus and are indicative of native ice composition, (2) "daughter" or "product" species produced by photolysis of gas-phase species in the coma used to determine parent species abundances, and (3) "distributed sources" or species whose distribution is incompatible with gas-phase chemistry and likely associated with photo or thermal degradation of refractories or volatile sublimation from icy grain in the coma. Ions observed in the coma are produced by ionization of neutral species and provide an indication of the parent species present in the nucleus.

Observations of primary species in the coma are straightforward and provide direct evidence that a species is present in the nucleus ice. Primary molecules $H_2O$, $CO$, $CO_2$, and HCN have been observed in multiple comets and the comae of at least three active Centaurs.

Uncertainties exist in determining the relative abundance of primary molecules in a coma based on their product species. The radical CN is a great example of this. During or shortly after an outburst, CN was observed around the Centaur Chiron when it was at a distance of ~11.26 AU from the Sun (Bus et al. 1991). A steady-state HCN outgassing rate of $3.7 \times 10^{25}$ $s^{-1}$ was estimated based on a total CN gas production rate of $5.3 \times 10^{29}$ $s^{-1}$. However, other observations of Chiron when it was closer to perihelion did not detect CN (Barucci et al., 1999; Rauer et al. 1997). Although HCN has been detected in comets as far out as 6.2 AU (Biver et al. 1998), its relatively high sublimation temperature might prevent sublimation at greater distances. It is therefore likely that the production of CN at large distances could be due to a combination of various volatile parent molecules, including HCN, but also small dust grains (composed of CHON) that can reach temperatures higher than the blackbody equilibrium temperature.

The radical CN has also been detected in the coma of 29P/S-W1(Cochran & Cochran 1991; Korsun et al. 2008; Ivanova al., 2016; 2018) with estimated production rates between $8.4 \times 10^{24}$ $s^{-1}$ and $3.34 \times 10^{25}$ $s^{-1}$. In this case it is likely a decay product of HCN, which has been detected at 29P/S-W1 and suggested to be produced by icy-grain sublimation (Bockelée-Morvan et al. 2022). The HCN abundance relative to water was a factor of 10 higher than values found in comets at 1 AU the Sun.

Distributed sources create further challenges because they influence the local composition of a coma such that it is different from the bulk coma composition. Extended sources can play an important role in coma composition for some small bodies, referred to as "hyperactive" comets, or comets whose active surface areas cannot account for their overall $H_2O$ production rate. The EPOXI mission to comet 103P/Hartley 2 revealed a complex coma environment. Strong $CO_2$ outgassing from the nucleus lobe region dragged water ice-coated grains into the coma that then sublimed and added to the overall gas content (A'Hearn et al. 2011) while $H_2O$ production in the nucleus waist region was dominated by direct release. Ground-based radio observations detected additional molecular contributions from distributed sources, including HCN and $CH_3OH$ (Drahus et al. 2012; Boissier et al. 2014). Recent work has demonstrated that icy grains are also a viable extended source for active Centaur comae with neutral gas such as HCN and $H_2O$ (Bockelee-Morvan et al. 2022).

The measurement of the $N_2$/CO ratio in comets plays a particularly important role in understanding planetesimal formation models and determining the solar nebula's physical properties during their creation. The condensation temperatures of $N_2$ and CO are much lower than most other species of volatiles in the PSN. Their presence in planetary building blocks provides indications of the formation temperature of these planetesimals and potentially the form of ice. Although laboratory experiments with amorphous ice suggest that CO and $N_2$ should be released simultaneously in the same proportion as they exist in the ice (Bar-Nun et al. 1988), some predict that dynamically new comets should have $N_2$/CO ratios in their coma similar to the solar valuethat decreases as comets are continuously exposed to solar radiation in the inner solar system due to the preferential loss of any $N_2$ in their outer layers (Owen and Bar-Nun 1995). Observations of this ratio in solar system small bodies are too few to make any firm conclusions and more work is needed (Anderson et al. 2022, 2023).

Because $N_2$ is difficult to detect through remote sensing, the only direct measurement of $N_2$ in a comet was made by the ROSINA instrument in comet 67P/C-G (Rubin et al. 2015), which allowed precise determination of $N_2$/CO of $5.7 \times 10^{-3}$. This ratio showed a significant depletion of $N_2$ compared to the proto-solar value of ~0.148, providing useful constraints on the formation temperature and ice type (Rubin et al. 2015; Mousis et al. 2016).

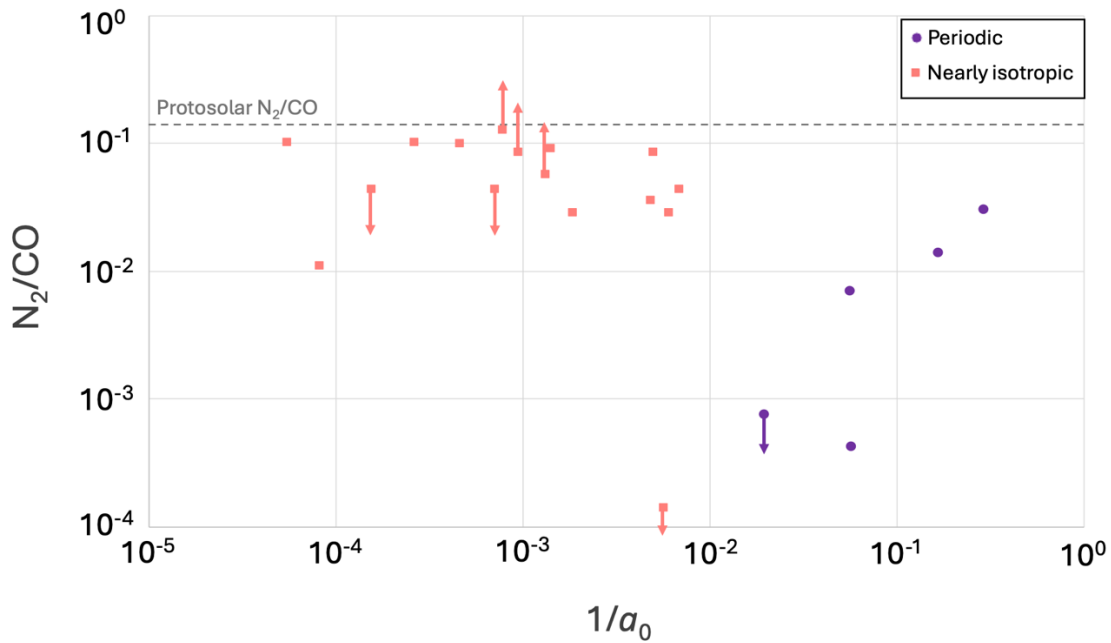

Figure 6.9: The ratio of $N_2^+$/$CO^+$ has been used as a proxy for determining the $N_2$/CO ratio in several comets. Compilation of estimated of $N_2$/CO for different comets and one Centaur depending on the bodies' semi-major axis show that the ratio may increase with increasing $1/a_0$ for periodic comets. Most of the nearly isotropic comets are closer to the solar value than periodic comets (Anderson et al., 2023). However, more observations are needed to determine if this is an artifact of observational limitations .

Ions are a useful tool for determining the abundance of species that are difficult to measure directly, specifically, CO and $N_2$ (Cochran, 2002; Fortenberry et al., 2021). The integrated intensities of the $CO^+$ and $N_2^+$ bands are typically used to estimate the ratio. Fig.

6.9 illustrates a comparison of the $N_2/CO$ derived from $N_2^+/CO^+$ for comets and one Centaur as a function of the semimajor axis, $a_0$. This shows that most nearly isotropic comets have ratios closer to the protosolar value than periodic coments as Owen and Bar-Nun (1995) predicted based on potential evolutionary processes that preferentially deplete the nucleus of $N_2$. The current data also potentially suggests an increase in $N_2/CO$ with increasing $1/a_0$ for periodic comets. However, Cochran et al. (2000) pointed out that a similar trend may only be partially objective, based on approximate estimates for some objects. Therefore, it is crucial to gather new observations and broaden the range of objects from different dynamical groups to draw reliable conclusions about determining the intrinsic values of $N_2/CO$ for them.

Furthermore, caution is needed in confirming detections of $N_2^+$. Low-resolution spectral observations have indicated the presence of $N_2^+$ emissions (see, for example, Wyckoff, 1989; Lutz et al., 1993; Korsun et al., 2006, 2008, 2014; Ivanova et al., 2016; 2018) but high-resolution spectral observations have not confirmed their presence (Cochran et al., 2000; Cochran, 2002). This could suggest that $N_2$ may in general be rare in comets, although it is important to note that increasing spectral resolution requires a narrow slit that may miss the ion tail. One exception is comet C/2016 R2 (Pan-STARRS), which exhibited strong $CO^+$ and $N_2^+$ emissions at a heliocentric distance beyond 3 AU (Cochran & McKay, 2018; Opitom et al., 2019; Anderson et al., 2022). The only Centaur where spectral emissions of $CO^+$ and $N_2^+$ have been detected is 29P/S-W1 (Fig. 6.10).

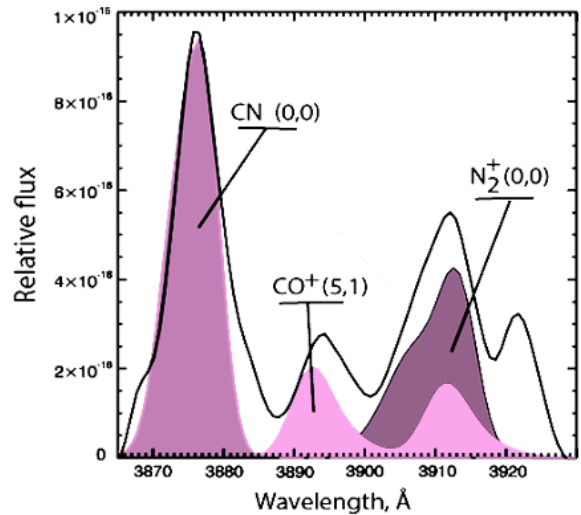

Figure 6.10: $N_2^+/CO^+$ ion ratios have been used as a proxy for the $N_2/CO$ ratio in 29P/S–W1. Observed (black solid curve) and calculated (filled) spectroscopic profiles of the $N_2^+$ (dark purple), CN (purple), and $CO^+$ (5, 1) band (lavender) for Centaur 29P/S–W1 (adapted from Korsun et al., 2008) reprinted with permission © Elsevier.

## 6.2.5 Current understanding of Centaur volatiles

We summarize below the state of knowledge for the surface and coma volatiles of seven Centaurs. They include active Centaurs 29P/S-W1, Echeclus, Chiron, (29981) 1999 TD10, and 39P/Oterma along with non-active Centaurs (10199) Chariklo and (5145) Pholus (hereafter TD10, Oterma, Chariklo, and Pholus).

**29P/Schwassmann-Wachmann 1.**

Centaur 29P/S-W1 is an active Centaur also classified as a JFC originating from the Kuiper Belt (Sarid et al., 2019). It has displayed significant activity since its discovery in 1927, remaining active throughout all observed time periods (Jewitt, 1990; Miles et al., 2016a,b,c; Whipple, 1980). Its orbit extends beyond Jupiter, so its continuous activity is not driven solely by water ice sublimation. However, the possibility of water ice evaporation cannot be entirely ruled out even at its location around 6 AU, as it can be stimulated by absorbing contaminants such as carbonaceous materials or silicates with a significant iron content (Hanner et al., 1981; Beer et al., 2006). The estimated radius of the 29P/S-W1 is ~30 km. This is significantly larger than known comets which have radii less than 10 km (Schambeau et al., 2015; Sierks et al., 2015; Kelley et al., 2017). The rotational period is still uncertain ranging from ~12 days based on coma features (Ivanova et al., 2012) to ~57 days based on suggested regularities in the outburst activity (Miles et al., 2016a), but these regularities have not been fully confirmed. In either case, the rotational period of 29P/S-W1 appears to be slower compared to other small bodies in the Solar System.

Because of the high activity levels, there are no surface observations. Observations of $H_2O$ in the coma are of interest. One study of a large outburst found that the estimated temperature at the subsolar point required 440% of the crystalline ice in the nucleus to be exposed and heated to explain the observed $H_2O$ production rate of $4.1 \times 10^{27}$ s$^{-1}$ (Bockelée-Morvan et al. 2022). This scenario is not physically possible. The small velocity offset observed for the $H_2O$ line indicated that the nucleus contributes minimally to water production, while an analysis of the temperature, velocity, and sublimation lifetime as a function of grain size revealed that long-lived icy grains with sizes exceeding a few micrometers played a significant role in the production of water molecules. Based on the water abundance observed in the coma, a lower limit of approximately $2 \times 10^8$ kg of icy grains was released during the outburst event (Bockelée-Morvan et al. 2022).

Observations of CO during this same outburst show a similar mass of this species released, indicating a substantial amount of material. As CO levels decreased, $H_2O$ levels exhibited a corresponding decrease. CO is often detected with production rates varying between $1 \times 10^{28}$ and $7 \times 10^{28}$ s$^{-1}$ (Pinto et al., 2022; Womack et al., 2017; Wierzchos and Womack, 2020). Further observations provide upper limits on abundance ratios of $CH_4$, $C_2H_6$, $CH_3OH$, and $H_2CO$ relative to CO that are consistent with the values in C/2016 R2 (PanSTARRS), a CO-rich Oort Clout Comet (OCC) (Roth et al. 2023). This is interpreted to suggest that CH$_4$ is preferentially stored with (polar) H$_2$O ice in the Centaur nucleus rather than with CO.

Table 6.1: Recent abundance measurements and upper limits for 29P/S-W1 shows a similarity in composition to CO-dominated comet C/2016 R2 rather than average comets.

| Ratio | 29P/S-W1 (%) | C/2016 R2 (%) | Avg. Comets (%) |
|---|---|---|---|
| CH$_4$/CO | < 0.98[a] | 0.59 ± 0.09[b] | 4.6–164[a] |
| C$_2$H$_6$/CO | < 9.1[a] | < 0.089[b] | 2.3–98[a] |
| CH$_3$OH/CO | < 21[a] | 1.04 ± 0.08[b] | 10–500[a] |
| H$_2$CO/CO | < 1.6[a] | 0.043 ± 0.006[b] | 1–81[a] |
| OCS/CO | < 22[a] | < 0.24[b] | 1.5–14[a] |
| N$_2$/CO | 0.014[c] | 0.089[c] | n/a |
| HCN/CO | 0.12 ± 0.03[d] | $(3.8 \pm 1.0) \times 10^{-3}$ [b] | 5.5 ± 1.4[b] |
| CO$_2$/CO | 4–18 x 10$^{-3}$ [e] | 0.05–0.16[f] | > 0.75[f] |
| Perihelion | 5.8 AU | 2.6 AU | n/a |

[a]Roth et al. 2023; [b]McKay et al. 2019; [c]Anderson et al. 2023; [d] Bockelée-Morvan et al. 2022; [e]Faggi et al. 2024; [f]Harrington Pinto et al. 2022.

As discussed earlier, HCN was observed in the coma of 29P/S-W1 and was found to be 10 times higher than comets at 1 AU (Bockelée-Morvan et al. 2022). Finally, observations have shown that the coma of 29P/S-W1 is CO$^+$ and N2$^+$-rich. The abundant presence of these highly volatile species could serve as drivers for its unique outburst behavior. However, there are no other large Centaurs similar to 29P/S-W1 within the orbits of Jupiter and Saturn, making it challenging to generalize its characteristics to other Centaurs.

**39P/Oterma**

39P/Oterma (39P) is an active Centaur, which has undergone major orbital changes in the last 90 years that had it fluctuating from a Centaur orbit to a closer-in Jupiter-family comet orbit and then back out to Centaur status. In 1943 Liisi Oterma discovered it with a visual magnitude of m 15 (Marsden 1962; Orchiston & Kronk 2009), when it had just passed perihelion (rhelio 3.4 au). It had often been reported with a non-stellar physical appearance, and in May 1943 it showed up on reflector plates as a star-like object surrounded by a small ring of nebulosity similar to a planetary nebula with a strong central condensation (Herbig & Mcmullin 1943). Since its orbit maintained a heliocentric distance between 3.4 and 4.5 au for several years, it was possible to observe it continuously with visual apparent magnitudes ranging from 15 to 19 from when it was discovered 1943 to 1961. When recovered in 2001, 39P appeared dimmer until 2023 with visual apparent magnitudes ranging from 22 to 24.

Prior to its discovery, 39P had a close approach to Jupiter in 1937 (MOID = 0.16 au), that shortened the orbital period from 18 years to 8 years. That close approach to Jupiter made the orbit transition from a Centaur orbit to a JFC orbit. Simulations of its orbit calculate that 39P had several close approaches to Jupiter and Saturn in recent centuries that eventually shortened the perihelion to 3.4 au which it maintained for about 30 years (Koon et al. 2000). From 1942 to 1962, 39P was dimmer than m ∼ 18 and occasionally exhibited a short tail (Marsden 1962). However, in its 1950 passage, 39P was seen to be its brightest with a m ∼ 14.5 (van Biesbroeck 1951; Orchiston & Kronk 2009). No mention of filters was made in the cited text, so we assume this is a visual magnitude. 39P was not seen again after 1962 until it was recovered in 2002 appearing to have weak/low activity and a brightness of r ∼22 mag (Jewitt 2009). It was lost to observers in 2002, and again recovered in 2019, showing what could be a compact coma and/or an elongated nucleus with r ∼24 mag (Schambeau et al. 2019).

The diameter of 39P is calculated to be 4–5 km. This is small compared to the 64 km diameter of 29P (Schambeau et al. 2021) but similar to Centaurs like 423P/Lemmon, P/2005 S2 (Skiff), P/2010 TO20 (LINEAR-Grauer), and 2000 GM137 that have diameters similar to the sizes of typical JFCs (Jewitt 2009; Schambeau et al. 2021, 2023). Photometry and image analysis from Gemini and the Lowell Discovery Telescope (LDT) indicated no extended emission. 39P's detections were consistent with that of a point source at the time of observations, thus suggesting that its bare nucleus was observed in the r′ filter. 39P's continuum shows absorption features at 2.0 and 3.1 μm attributed to water ice. From the ice absorption features, we can see that the band depths for 39P closely match 103P, except where gas emission bands are present. This suggests that the water ice grain size of 39P is similar to what Protopapa et al. (2014) measured in the coma of 103P: about 1 μm. This small grain size supports that the water ice being released from 39P and observed by JWST is in the coma of 39P rather than a detection of its surface (Protopapa et al. 2018). However, a definitive conclusion cannot be made, given that the spatial profile of the continuum lacks a significant extended source. As an alternative, the surface of 39P may be covered in widespread micrometer-sized ice grains. We also see that the continuum of 39P, 67P, and 103P are in excellent agreement from 1.0 to 1.9 μm and at 3-μm all three objects show an absorption feature.

39P is the first centaur to have $CO_2$ detected (Harrington Pinto et al. 2022, 2023). Prior to this $CO_2$ had rarely been detected in comets, and never in Centaurs. JWST observations provide a $CO_2$ detection with a production rate, $Q$, of $(5.96 \pm 0.80) \times 10^{23}$ molecules s$^{-1}$. This is the lowest detection of $CO_2$ yet for any comet or Centaur, and it is only possible due to the capabilities of JWST NIRSPEC. While CO and $H_2O$ were not detected, the 3-sigma upper limits of $Q_{CO} < 12.1 \times 10^{23}$ molecules s$^{-1}$ and $Q_{H_2O} < 10.0 \times 10^{23}$ molecules s$^{-1}$. The $CO_2$/CO ratio for Oterma is >0.49 (Harrington Pinto et al. 2023), which is much larger than the ratio observed for 29P/S-W. These differences highlight an important point that the composition of Centaurs could be very diverse. This large difference in CO and $CO_2$ abundances could be due to the different thermal heating that each of these objects have undergone. CO is more volatile than $CO_2$, so since CO is lower in 39P than 29P it could indicate that 29P is more pristine. Adding to the mystery, while 39P has traveled closer to the Sun than 29P, 29P is experiencing solar heating more consistently than 39P which makes it puzzling how 29P is still actively outgassing CO.

Centaurs are active while still so far from the Sun. Observations of their dust and gas comae can give insight into evolutionary processes and serve as powerful tools for tracing primitive material from the formation of the solar system (Tegler et al. 2003; Lisse et al.

2022). JWST gives us the first opportunity to measure the volatile content in many of these distant bodies before they are significantly processed by the Sun.

**(60558) 174P/Echeclus.**

Echeclus is an active Centaur demonstrating outburst activity that does not depend on its location over its orbit and at distances as great as 13 AU. It was initially discovered on March 3, 2000, has a diameter of 64.6±1.6 km, and an albedo of around 5 percent (Duffard et al., 2014). Spectral absorption features of water ice have not been observed on its surface (Seccull et al., 2019). It has a steeper red color in visible wavelengths compared to near-infrared wavelengths. Preliminary modeling for analyzing the average polarization values indicates a surface mixture of carbon, silicates, and some ice – consistent with Rosetta measurements of 67P/C-G. However, the presence of ice is suggested to be only a few percent, significantly lower than the few tens of percent observed in distant comets.

CO emission was detected in the coma of Echeclus at 6 AU (Wierzchos et al., 2017; Kareta et al., 2019) with an estimated production rate of $7.7 \times 10^{26}$ $s^{-1}$. Echeclus has been observed to have a blue coma (Seccull et al. 2019), similar to observations of the CO-rich comet C/2016 R2 (Biver et al. 2019).

**(2060) Chiron.**

Chiron is an active Centaur that was discovered on November 1, 1977 demonstrating the recurrent occurrence of comet-like activity (Duffard et al., 2002). Chiron's mean radius is estimated to be approximately 107.8±4.95 km, with an albedo of 0.08 (Fornasier et al., 2013). Detailed analysis of its light curve has yielded a well-defined rotational period of 5.918 hours with a minor brightness variation ranging from 0.05 to 0.09 magnitudes suggesting a predominantly spheroidal shape (Marciali and Buratti, 1993) and possible surface volatile activity. Furthermore, evidence suggests the possible presence of rings (Ortiz Moreno et al., 2015).

Spectral observations have shown water absorption features on the surface, but the data quality is insufficient to definitively determine if the ice is crystalline or amorphous (Foster et al., 1999). The surface reflectivity gradient appears to more closely resemble that of C-type asteroids than that of cometary nuclei. Small variations in the optical reflectivity gradient correlating with episodes of activity may be attributed to fluctuations in dust production. Long-term variations in brightness indicate surface activity and the existence of volatile substances. The polarization phase behavior of Chiron in its non-active period is significantly different from any other Solar system bodies studied so far (Bagnulo et al. 2006).

CO has been observed in the coma at a production rate of $1.3 \times 10^{27}$ $s^{-1}$ (Womack & Stern 1999). This CO may have resulted from an outburst, although the production rate is also sufficient to drive the observed dust coma activity. As noted earlier, CN was observed in the coma when Chiron was at a distance of ~11.26 AU from the Sun (Bus et al. 1991) but not when it was closer to perihelion (Barucii et al., 1999; Rauer et al. 1997) suggesting that HCN may not be the main source of this radical.

**(29981) 1999 TD10**.

This active Centaur has a perihelion distance greater than 12 AU, a diameter of 103.7±13.5 km (Barkume et al., 2008), and a period of 15.448 ± 0.012 hours (Mueller et al., 2004; Choi et al., 2003; Rousselot et al., 2005). It exhibits weak activity (Choi et al., 2003). The normalized reflectivity gradient on the surface is similar to the median value observed for Centaurs (Mueller et al., 2004), and no water has been detected in surface spectral observations (Barkume et al., 2008).

**(10199) Chariklo.**

Chariklo is a non-active Centaur that was discovered on February 15, 1997 and has a diameter of approximately 248 ± 18 km (Fornasier et al., 2013), ranking among the largest known Centaurs. Its rotation period is ~7 hours. Photometric analysis of observations has unveiled the presence of two distinct narrow rings (Braga-Ribas et al., 2014). Infrared observations of Chariklo's surface have periodically indicated the presence of water ice (Fornasier et al., 2013 and references therein), which may be located within its rings. It is not clear if the water ice is amorphous or crystalline (Dotto et al., 2003). Polarimetric observations revealed evidence of surface heterogeneity (Belskaya et al., 2010).

**(5145) Pholus.**

This non-active Centaur is one of the most primitive objects known within its class (Cruikshank et al., 1998), with a diameter of approximately 107 ± 19 km and an albedo of 0.12 (Duffard et al., 2014). Its rotation period is 9.98 ± 0.02 hours (Tegler et al. 2005, Buie and Bus 1992, Farnham 2001). The surface displays a highly conspicuous reddish hue (Cruikshank et al., 1998; Fornasier et al., 2009) and the reflectance spectrum reveals the presence of two distinct components that are spatially separated. These components include dark amorphous carbon and an intimate mixture of water ice, methanol ice, olivine grains, and complex organic compounds known as tholins. Polarimetric observations have unveiled a significant negative polarization at specific phase angles that is distinctly different from that observed in TNOs (Belskaya et al., 2010).

## 6.3 Centaurs in solar system context

Centaurs play an important role in understanding the formation and evolution of the solar system. Their volatile composition can tell us where they formed and how they relate to other small body populations, providing critical pieces of a dynamical puzzle that is difficult to constrain with the current data available. We review the current state of knowledge for the volatile composition of other small body populations providing a baseline for future comparisons as more data becomes available for Centaurs.

## 6.3.1 Census of observed small body volatiles

**Oort Cloud and Jupiter Family Comets:**

In general, comets originating from the Oort Cloud are called OCCs while ones that originated as TNOs are referred to as JFCs. However, there are exceptions where comets originating from the Oort Cloud are captured into shorter orbits causing them to exhibit similar orbital characteristics to JFCs. Chemical diversity is evident among comets both in parent volatiles and daughter species. A review of the average mixing ratios of eight parent volatiles in 30 comets ($CH_3OH$, $HCN$, $NH_3$, $H_2CO$, $C_2H_2$, $C_2H_6$, $CH_4$, $CO$) suggest that OCCs are volatile-rich (Fig. 6.11). This review also proposed an overall depletion of these volatiles in JFCs compared to OCCs that was most pronounced for the four species of highest volatility ($C_2H_2$, $C_2H_6$, $CH_4$, and $CO$), suggesting that thermal evolution may affect the compositional differences between the populations (Dello Russo et al. 2016). There were weak positive correlations between $NH_3$ and $HCN$, and between $NH_3$ and $H_2CO$ in both OCCs and JFCs and strong correlations between $HCN$, $C_2H_2$, and $C_2H_6$ (Dello Russo et al., 2016). However, it should be noted that more recent work with improved statistics indicates broader compositional diversity in each dynamical class, and found that with the exception of CO (for which JFCs are on average depleted compared to OCCs) there are no significant compositional differences between parent volatile abundances among the dynamical families (Biver et al. in press). This suggests that many proposed differences may be related to sample size biases.

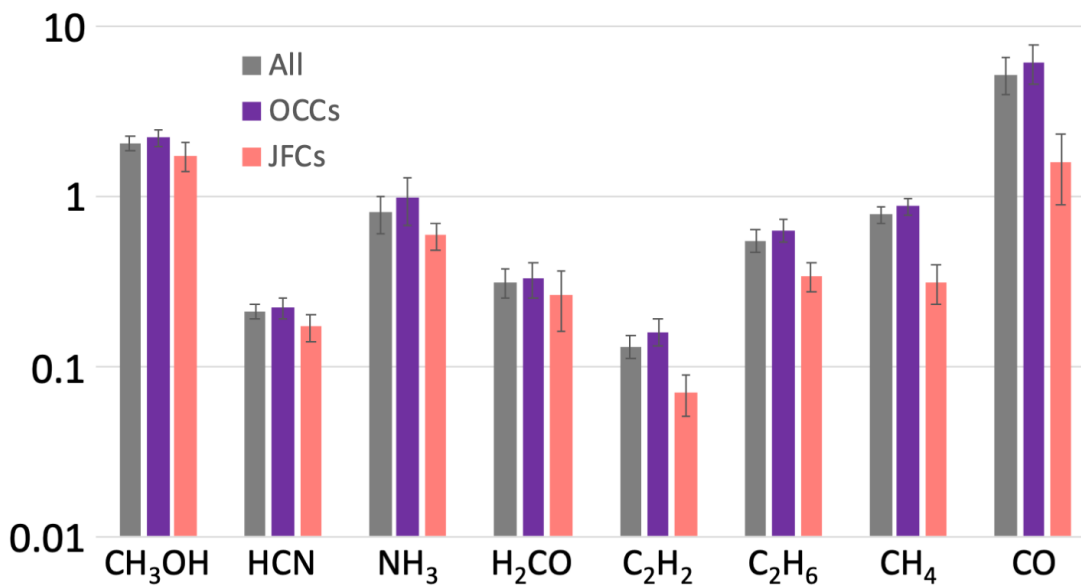

Figure 6.11: The relative abundance of different volatile species in comet families can help to constrain their origins. Average mixing ratios (% with respect to H2O) for up to 27 comets based on availability of measurements compared to the mixing ratios for the OCCs (up to 19) and JFCs (up to 8) in the sample (adopted from data by Dello Russo et al. 2016).

The presence of hypervolatiles in OCCs can help to constrain models to determine whether Oort Cloud objects were ejected to this region by giant planet migration early (<20 million years) or later (0.05–2.0 billion years). Dynamical models suggest that hypervolatile-rich planetesimals in the Oort Cloud may have been placed within approximately 20 million years of solar system formation and represent the earliest objects in the Oort Cloud providing potential insights into $CO/N_2/CH_4$ ratios in the protoplanetary disk (Davidsson et al. 2021; Prialnik et al. 2021; Steckloff et al. 2021).

Research on radicals (OH, CN, $C_2$, $C_3$, NH) in 85 comets indicated that about half of JFCs are classified as carbon-chain depleted, a much larger percent than the observed OCCs (A'Hearn et al., 1995). This is interpreted to be caused by differences in carbon-chain chemistries in the initial conditions where JFCs formed and not caused by their extensive thermal processing due to numerous perihelion passages (A'Hearn et al., 1995; Fink, 2009). If this is the case, it could also imply that pre-cometary ices that formed JFCs were subjected to conditions that, on average, selectively depleted the most volatile species.

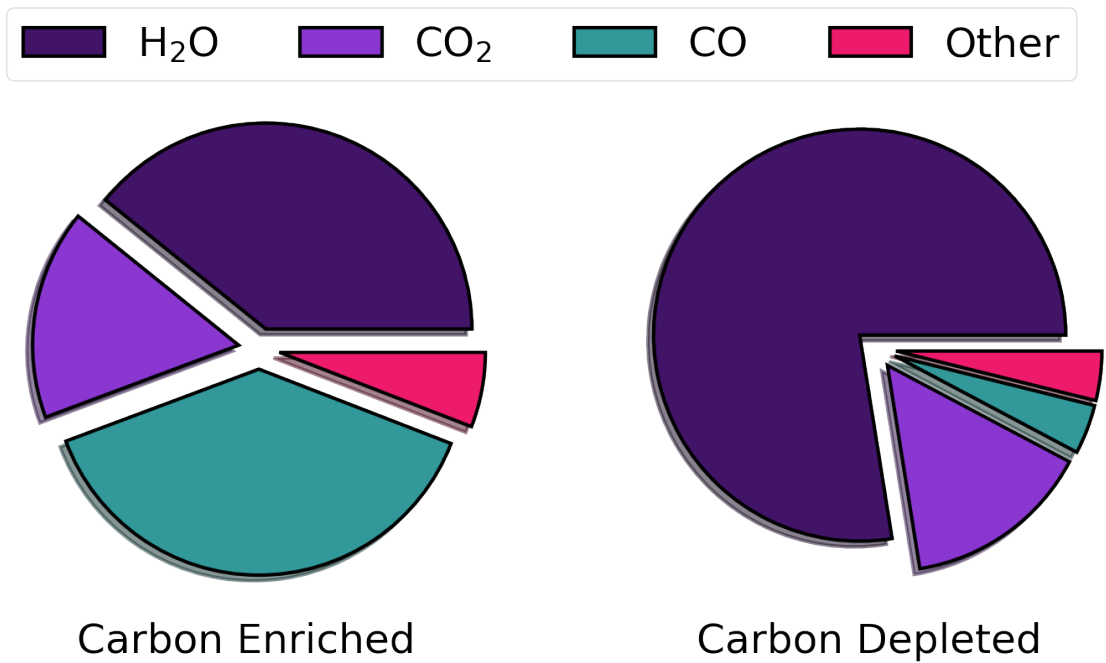

Figure 6.12: Typical compositions of carbon-enriched and -depleted comets illustrate two compositional families of comets that may indicate different formation regions. The carbon enriched comet is C/2006 W3 Christensen (Ootsubo et al 2012). The carbon depleted pie chart is representative of most comets for which production rate measurements of $CO_2$, CO and $H_2O$ exist. This is a generalized version of an analogous figure in McKay et al. 2019 (Seligman & Moro-Martín 2022), reproduced with permission © Taylor & Francis.

However, our understanding of comet composition and how it varies from one comet to the next remains limited. In general, we have found that the observed volatiles in typical

OCCs and JFCs primarily consist of $H_2O$ ice. $CO_2$ and CO ice are the next two most abundant volatiles followed by carbon- and nitrogen- bearing species such as $CH_4$, $C_2H_2$, $C_2H_6$, $CH_3OH$, $NH_3$ and HCN (Dello Russo et al. 2016; Biver et al. 2023). The discovery of C/2006 W3 Christensen followed by C/2016 R2 (Pan-STARRS) indicated the potential existence of a different class of comets in the solar system based on composition and described as carbon-rich. The composition of C/2006 W3 Christensen compared to the average composition of JFC and OCC comets is illustrated in Figure 6.12. Carbon-enriched comets are proposed to have formed in the region of (Mousis et al. 2021), at (Price et al. 2021), or outside of the CO ice-line in the PSN (Seligman et al. 2022).

Recent work has compared the composition of 29P/S-W1 during an outburst against parent volatiles measured in comets at radio and near-infrared wavelengths (Roth et al. 2023, PSJ, 4, 172). Stringent upper limits on abundance ratios of hypervolatiles ($CH_4$/CO and $C_2H_6$/CO) and hydrocarbons ($CH_3OH$/CO and $H_2CO$/CO) were most consistent with values measured in C/2016 R2 (PanSTARRS). However, compositional comparisons between Centaurs and comets are challenging, as the vast majority of parent volatile studies in comets take place in the inner solar system (heliocentric distances $< \sim 2$ au) where $H_2O$ is vigorously subliming, in contrast to the much larger heliocentric distances of Centaurs.

**Active TNOs**

TNOs are categorized into two main groups: the classical and resonant entities of the Kuiper belt and the scattered disc and detached objects, with the sednoids representing the most distantly located members. Observations of activity for distant TNOs are limited and direct observation of sublimation of gases is limited to Pluto (Gladstone & Young 2019; Young et al. 2020). Therefore, additional mechanisms are required to explain activity at significant distances from the Sun. What is notable is that many distant comets exhibit long duration activity that cannot be explained by crystallization of amorphous water ice, so further studies are needed.

Dust production in these distant comets is significantly higher than that of short-period comets, similar to new comets entering the inner solar system for the first time. It is generally higher before perihelion passage than after, and appears to contain relatively large particles with an icy component (Korsun et al., 2006; 2010; 2014) with sublimation times lasting several years. The structures in distant comet comae do not resemble the comet tails that form at close heliocentric distances. They do not typically have an internal structure, display an almost constant width along the tail, and are often strongly bent. Some distant comets have asymmetric and elongated comae, without tails.

Polarimetric studies suggest that the composition of particles in distant comets differs from the composition of dust in short-period comets (Dlugach et al., 2018; Ivanova et al., 2015a, 2015b; 2019b; 2021; 2023). For example the coma of C/2014 A4 (SONEAR) is dominated by submicron particles composed of a large amount of ice and tholin-like organic substances (Ivanova et al., 2019b) while the coma of C/2011 KP36 (Spacewatch) is formed by particles of various sizes consisting of water ice, $CO_2$ ice, and refractory material (Ivanova et al., 2021).

Spectral observations of comet C/2002 VQ94 (LINEAR) when it was at 7.33 AU from the Sun revealed the presence of $CO^+$ and $N_2^+$ emissions (Korsun et al., 2006; 2008) interpreted to mean that this comet is enriched in CO and $N_2$. The emissions disappeared when the comet was at a distance of 9.86 AU.

**Active asteroids**

The main asteroid belt contains rare asteroids that are classified as active asteroids based on the Tisserand parameter, the presence of a coma, and the detection of water and other volatiles (Jewitt et al. 2015). In these small bodies, manifestations of activity can be more diverse than comets. Similar to comets, active asteroids shed a noticeable amount of material, with activity varying from short-lived events (Neslusan et al. 2016; Ivanova et al. 2023), to recurring activities, exemplified by objects like 133P/Elst-Pizarro (Hsieh et al., 2010; Jewitt et al., 2014) and 6478 Gault (Chandler et al 2019; Ivanova et al., 2020).

Several mechanisms that can trigger such activity have been proposed including ice sublimation with associated dust expulsion, rotational breakup, impact events leading to dust ejection, thermal fractures, and rotational fission of contact binary asteroids. However, it is easier to rule out specific processes than to identify a definitive cause. This is partly due to limitations in data, but it also reflects the intricate nature of the observed activity. For example, in cases where sublimation is believed to be the primary driver of activity, an impact or other disruptive event may be required to expose buried ice and trigger sublimation, with rapid rotation potentially aiding mass loss.

Active asteroids demonstrate remarkable orbital stability over extremely long timescales, often exceeding 100 million years (Jewitt 2009; Hsieh et al., 2012, Stevenson et al., 2012). However, exceptions like 238P and 259P have been identified as unstable over shorter timescales, around 10 million years. Several active asteroids have been associated with collisional asteroid families and clusters, which could play a role in preserving ice over long timescales through shielding within a larger parent body until more recent exposure to direct solar heating triggers activity.

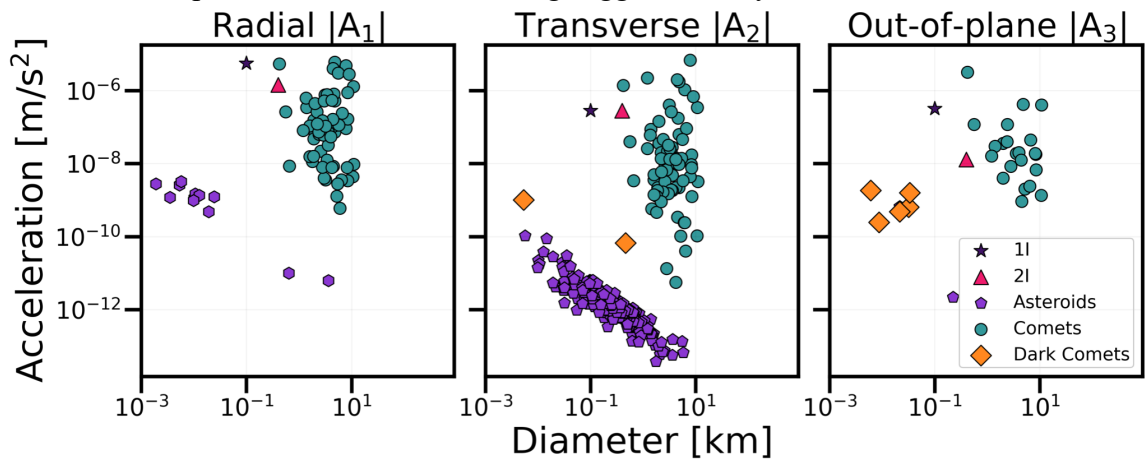

Figure 6.13: A new class of potentially active asteroids, referred to as dark comets, are photometrically inactive but exhibit significant nongravitational accelerations. Measured components of nongravitational accelerations of small bodies with known diameters in the radial A1 direction (left), transverse A2 direction (middle) and out-of-plane A3 direction (right). (Seligman & Moro-Martin 2022), reproduced with permission © Taylor & Francis.

Recently, a new class of potentially active asteroids was also discovered (see Fig. 6.13), the so-called "dark comets" and include (523599) 2003 RM, 1998 KY26, 2005 VL1, 2016 NJ33, 2010 VL65, 2006 RH120, and 2010 RF12. These dark comets are photometrically inactive asteroids that exhibit significant nongravitational accelerations that are (1) too strong in magnitude and (2) in the wrong direction to be driven by typical radiation based effects seen on asteroids, such as the Yarkovsky effect or radiation pressure (Seligman & Moro-Martin 2022). Because these objects are photometrically inactive, it is unclear whether the nongravitational accelerations are driven by outgassing. However, to date there is no alternative mechanism known that could explain the observed nongravitational accelerations.

If these objects are actively outgassing, it is possible that they were delivered to their current orbits from the Jupiter family comets, and presumably from the Centaur region prior to that. Therefore, it is possible that these dark comets represent one possible end state of Centaur evolution into the inner Solar System. Future observations will be required to measure the outgassing volatile content of these objects, and comparisons to JFCs and Centaurs will likely be revealing if these objects share similar evolutionary histories. Excitingly, one of the dark comet candidates, 1998 KY26, is already the target for the extended Hayabusa2 mission and exhibits favorable viewing geometry before 2025. The Legacy Survey of Space and Time (LSST) planned to be conducted with the forthcoming Vera Rubin Observatory is poised to further transform our understanding of these classes of objects.

**Interstellar comets**

We have limited knowledge of the census of volatiles on interstellar comets, given only two known objects. The first interstellar object 1I/'Oumuamua displayed no detectable cometary activity either in volatiles or dust reflectance but exhibited nongravitational acceleration (Micheli 2018). The mechanisms proposed include outgassing invoked compositions of $H_2$, $N_2$, CO or $H_2O$ (Seligman & Laughlin 2020, Desch & Jackson 2021, Jackson & Desch 2021). The interstellar comet 2I/Borisov, which had a closest approach to the Sun of ~2 AU, displayed a significant and distinct cometary tail and was enriched in hypervolatiles compared to $H_2O$ (Cordiner et al. 2020, Bodewits et al 2020).

Given our limited knowledge of volatiles in interstellar comets due to our limited detection of objects in the population, it is challenging to draw population level comparisons with well characterized solar system small body populations. However, 2I/Borisov was enriched in hypervolatiles to a higher extent than observed in Centaurs and it could be explained as having formed at further stellocentric distances in its host disk. It is proposed to have formed at or exterior to the CO snowline, possibly in an M-dwarf system where a larger fraction of the total area in the protostellar disk would be exterior to the CO snowline (Cordiner et al 2020). However, observations of Centaurs are typically taken at much larger heliocentric distances (>3 AU) than the observations that were taken of 2I/Borisov (2 – 2.7 AU), so it is likely that some volatile reservoirs ($H_2O$) in Centaurs that have been observed were not active during the observations.

### 6.3.2 Comparison to solar system models

A central goal in the study of primitive small bodies is using their dynamical history and composition to trace back to the chemistry and physics present in the protoplanetary disk midplane at the time of planet formation. Such an analysis requires a statistically significant sample, which has historically been difficult to obtain for volatile composition studies of Centaurs based on the few active Centaurs available and observational challenges associated with the limitations of ground-based facilities.

Overall the current state of knowledge of primordial solar system volatiles is limited. Fig. 6.14 illustrates the current state of knowledge for the bulk composition of volatile elements in the giant planet atmospheres and known small body populations compared to the composition of the PSN (Lodders 2021). From this figure we can see that Jupiter is the only giant planet with a complete set of observations thanks to the in situ measurements made by the Galileo Probe Mass Spectrometer (GPMS; Mahaffy et al. 2000; Wong et al. 2004) and the Juno remote observations with the Microwave Radiometer (MWR; Li et al. 2017, 2020).

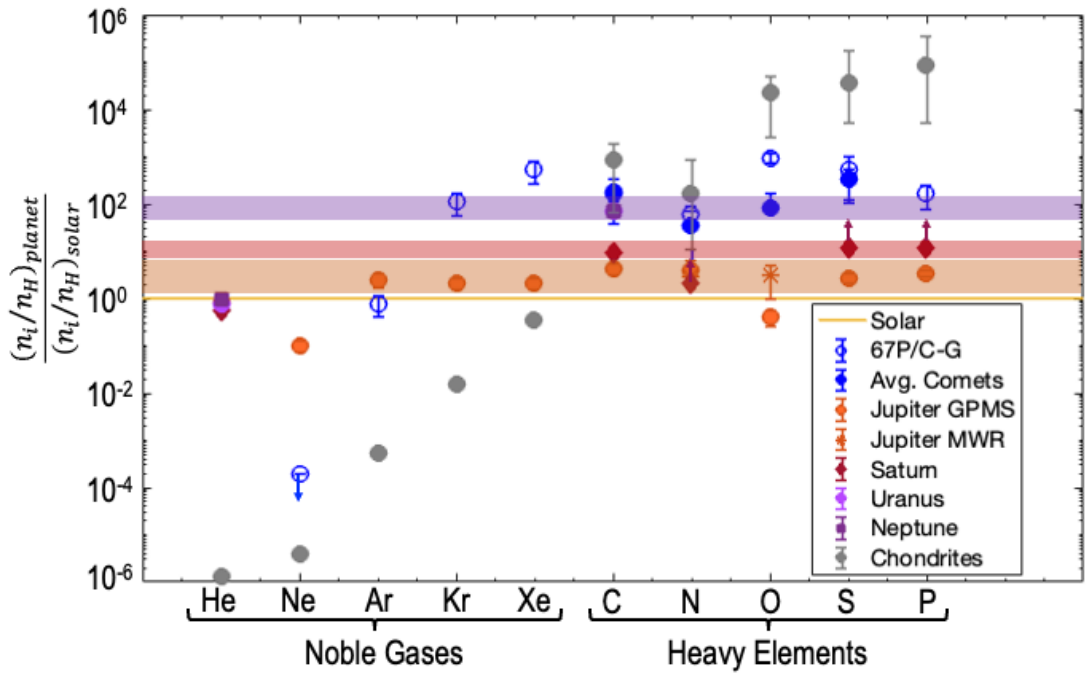

Figure 6.14: Available observations of the heavy element and noble gas abundances in the atmospheres of the four giant planets relative to PSN composition along with analogs for planetesimals from the time of solar system formation can help to constrain the building blocks for the giant planets. The color of the points indicate the origin of the measurements, and references are provided in the main text.

Bulk composition constraints for small bodies are currently limited to chondrites (Lodders, 2021) and cometary ices (Le Roy et al. 2015; Rubin et al. 2019) and show significant fractionation of the elements compared to solar values. The heavy noble gas relative abundances are particularly interesting, as shown in Fig. 6.15 where solid materials differ by orders of magnitude compared to the solar values. Chondrites found in meteorites that impact the Earth provide an analogue for the volatiles contained in asteroids and the refractory material of icy bodies like comets. The comet observations shown in Fig. 6.14 are based on icy materials observed in the comae of comets that either originate as TNOs (JFCs) or from the Oort Cloud (OCCs). Centaurs are intermediate in distance from the Sun between asteroids and TNOs, residing within the orbits of the giant planets.

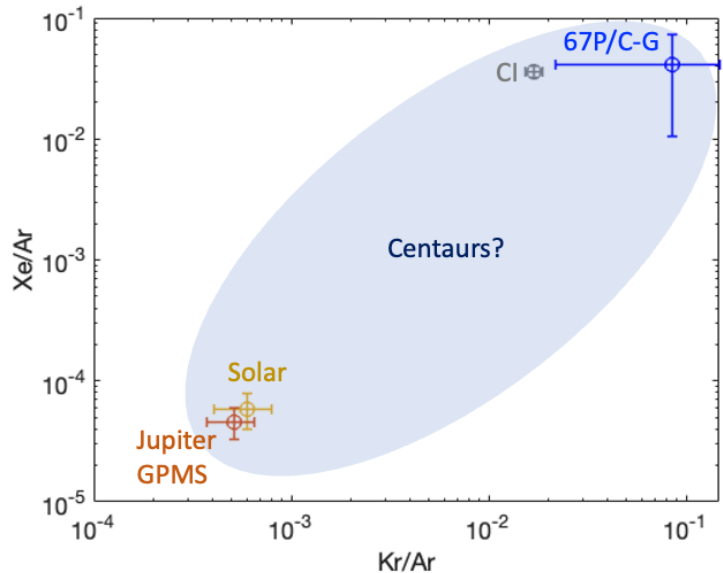

Figure 6.15: Available relative noble gas abundances in Jupiter's atmosphere, the Sun, and the limited knowledge for analogs for solid building block materials. Adding more information on the noble gas abundances can help us to understand when and where each giant planet formed and how they migrated after formation.

Fig. 6.16 illustrates the current state of knowledge for three isotope ratios of interest within various solar system bodies. Carbon isotope ratios, $^{12}C/^{13}C$, are limited to a narrow range around a value of ~90 with error bars representing a range between 65 and 110 (e.g. Alexander et al. 2013; Mandt et al. 2009). The one exception is the Saturnian satellite Phoebe which is highly enriched in the heavy isotope compared to any other solar system body (Clark et al. 2019).

The D/H ratio in water is a useful proxy for the formation location of solid building blocks because D/H in water varied as a function of distance from the Sun as the solar system formed, increasing in value with greater distance as the temperature decreased (Aikawa & Herbst, 1999). As Fig. 6.16 illustrates water incorporated into chondritic material (Alexander et al. 2012) likely formed in warmer conditions than OCCs (Biver et al. 2006; 2016; Bockelée-Morvan et al. 1998; 2012; Brown et al. 2012; Gibb et al. 2012; Hutsemekers et al. 2008; Lis et al. 2019; Meier et al. 1998; Paganini et al. 2017; Villanueva et al. 2009; Weaver et al. 2008). The D/H in JFCs suggested that this class of comets formed in similar conditions to chondrites (Hartogh et al. 2011; Lis et al. 2013; 2019). Initial reports of one of the highest reported D/H values in the Rosetta JFC target comet, 67P/C-G appeared to contradict this possibility (Altwegg et al. 2015). However, a recent reanalysis of the Rosetta observations making use of over 4000 data points from the mission discovered that the D/H in coma near the spacecraft location was highly influenced by enriched ice sublimating from dust grains (Mandt et al. 2024). When accounting for the role of dust in the measurements, the D/H was revised to a lower value that is much closer to chondrites (Mandt et al. 2024).

Three moons of Saturn are also illustrated. Titan's observation is in methane, which is different from the D/H in water on Enceladus (not shown; Waite et al. 2009) and Iapetus suggesting that Titan's methane is primordial and not produced from interacting with water through hydrothermal chemistry in Titan's interior (Mousis et al. 2009). As with the carbon isotopes, Saturn's moon Phoebe is enriched in the heavy isotope far more than any other solar system body (Clark et al. 2019).

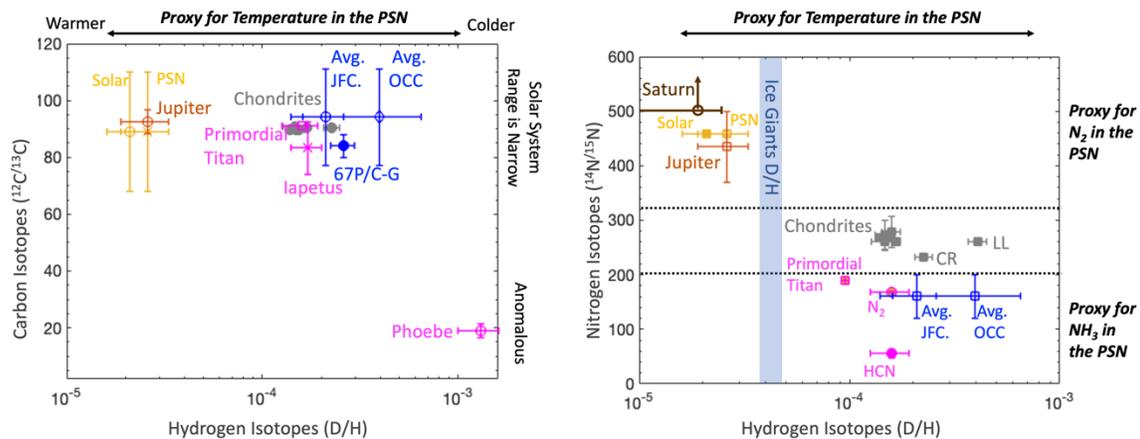

Figure 6.16: Current state of knowledge for isotope ratios that can be used to determine the origin and evolution of materials throughout the solar system (see text for individual measurement references).

Finally, the nitrogen isotope ratio, $^{14}N/^{15}N$, is a useful tracer for the source of nitrogen in the PSN. The bulk of nitrogen in the PSN was in the form of $N_2$, which had a high ratio. Other species, including organics found in chondrites as well as $NH_3$ and HCN ices were more enriched in the heavy isotope.

At the present time insufficient information is available about the bulk composition of Centaurs to include them in Fig. 6.14 and no noble gas abundances or isotope ratios have been measured in Centaurs. These measurements are needed because as likely dynamical progenitors to JFCs, Centaurs may provide a window into TNO material that has undergone less potential thermal processing than that preserved in short period comets. Although earlier studies proposed that OCCs formed between heliocentric distances ~5 – 30 AU and JFCs formed at larger heliocentric distances, the presence of crystalline silicates in some comets require that processed material from the hot inner disk was incorporated into their nuclei along with hypervolatiles from the cold outer disk. This suggests a more "spatially mixed" comet formation region.

Furthermore, the composition of ices preserved in Centaurs can improve comparisons with ice-phase astrochemical models of protoplanetary disk midplanes. Recent work comparing the composition of JFCs and OCCs measured to date found that no single time or location in the midplane could fully account for the molecular diversity measured in comets (Willacy et al. 2022). As volatile inventories for Centaurs continue to improve with the latest ground- and space-based facilities, similar comparisons between the primitive small bodies and midplane ice-phase chemistry will become possible.

## 6.4 Future prospects

Volatiles in Centaurs provide important clues to the conditions in which they formed and can help us to better understand the bigger picture of solar system formation. Furthermore, the form of volatiles may play an important role in Centaur activity, more so than regular comets where activity is driven by sublimation. We close this chapter by outlining the most important measurements needed through future exploration of Centaur volatiles and describe the future prospects based on upcoming capabilities.

### 6.4.1 Open questions

We need to learn more about the composition and form of ice in Centaurs compared to other small bodies in order to understand their role in the history of the solar system. This information will provide vital clues about the formation conditions of small bodies and how they may have contributed to the formation of the solar system.

The summary of observed volatile elements that can be used to trace formation and evolution processes for the solar system illustrated in Fig. 6.14 is clearly incomplete. More definitive constraints on the abundances of carbon-, nitrogen-, and sulfur-bearing volatiles relative to water in Centaurs are needed to place them in the context of comets and asteroids. Additionally, an improved understanding of comet composition is needed. Most comet measurements are from a single point in a comet's orbit, and it is not clear if these observations are reflective of the bulk composition. Measuring the composition of the comae of several comets over a range of distances from the Sun is needed to resolve this question.

Additionally, measurements of noble gases are needed in Centaurs and in more comets. Noble gases are particularly useful for tracing the sources of volatiles because they are nonreactive and not influenced by chemistry. As Fig. 6.15 showed, the relative abundances of the heavy noble gases can separate solid materials from the bulk composition of the PSN and we need to determine where Centaurs fit in. It should be noted that because noble gases are non-reactive they cannot be measured remotely and require spacecraft observations in situ. Opportunities to determine noble gases should be prioritized in planning future missions to Centaurs.

Finally, measurements of stable isotope ratios in Centaur volatiles are needed. The D/H ratio in water will provide clues to the temperature conditions where the water ice formed. The nitrogen isotopes can provide indications of the source of nitrogen and the temperature conditions for formation while carbon isotopes can provide constraints on volatile evolution.

## 6.4.2 Recommended activities

A combination of archival studies to search for activity not previously detected in known Centaurs, Earth-based observations from the ground and space, as well as spacecraft missions either to fly by or orbit Centaurs is needed to obtain the observations described above.

**Ground-based**

Ground-based observations of Centaurs, facilitated by utilizing surveys with large telescopes such as the LSST, can play a pivotal role in obtaining many of the needed observations. Large telescopes are particularly important because of the inherently low brightness of Centaurs. A quasi-simultaneous approach to ground-based observations that combine photometry, spectroscopy, and polarization studies are of high value and offer a comprehensive view of Centaurs. This approach has proven effective for several distant comets (Ivanova et al., 2019b, 2021, 2023).

Furthermore, long-term monitoring of Centaurs is indispensable for uncovering variations with time and heliocentric distances. These variations hold the key to understanding the surface dynamics and composition of both non-active and active Centaurs. Through extended observation campaigns, astronomers can capture the subtle changes that occur on these distant objects over time, shedding light on their non-typical behavior and evolutionary processes.

**Space-based from Earth**

Observations with the recently launched James Webb Space Telescope are already taking place and revolutionizing our understanding of the distant universe and our solar system. The infrared capabilities of JWST allows us to characterize the volatile content of active Centaurs. Prior to JWST, 29P/S-W1 was the only Centaur with measured CO production rates and upper limits on $H_2O$ and $CO_2$ was (see section 6.3.1). So far, JWST Cycle 1 observations have included observations of 29P/S-W1, 39P/Oterma, C/2004 A1 (LONEOS), C/2008 CL94 (Lemmon), C/2014 OG392 (PanSTARRS), and C/2019 LD5 (ATLAS). Successful detection of $CO_2$ gas emission by 39P/Oterma and upper limits of CO and $H_2O$ provide limits on the mixing ratios of $CO/CO_2 \leq 2.03$ and $CO_2/H_2O \geq 0.60$ (Pinto et al 2023). Moreover, JWST imaging provides constraints on the radius of the nucleus between 2.21 to 2.49 km. Further observations of 29P/S-W1 revealed detections of CO and $CO_2$ isotopologues, $^{13}C^{16}O_2$, $^{13}C^{16}O$, $^{12}C^{17}O$, and $^{12}C^{18}O$, marking their first detection in a Centaur (Faggi et al. 2023). While these are some of the first measurements of active Centaurs with JWST, they demonstrate how characterization of volatile content of Centaurs in the near term future will greatly advance with the capabilities of JWST.

The forthcoming Habitable Worlds Observatory (HWO; NASEM, 2020) poses significant prospects to advance our understanding of the formation and evolution of the Solar System via Centaur science. HWO is recommended to have ultraviolet spectroscopic capabilities that would enable measurements of the production rates in active Centaurs of nitrogen-bearing species such as $N_2$, $NH_3$ and HCN, nitrogen isotopes, sulfur bearing species and isotopes, and oxygen isotopes. It is also possible that this observatory could measure D/H in surface water ice or water released from grains in the coma, which would provide information into the still open question of volatile delivery to the terrestrial planets.

**Centaur spacecraft missions**

Spacecraft missions, whether flying by a Centaur or orbiting one, would provide access to information that is not available from a distance. A flyby of a Centaur would provide an opportunity to image the surface and make spectroscopic observations of surface ices. Thermal imaging of the nucleus could provide important constraints on the conditions for activity if the Centaur is observed to be active. Mass spectrometers capable of measuring the composition of ions and neutral particles could detect or set upper limits for activity during the flyby and measure the composition of volatiles being released if the Centaur is active. Dust detectors could provide composition measurements for any dust being released. An orbital mission would provide even more information through detailed mapping of the surface geology and composition. Furthermore, ongoing monitoring for activity and characterization of any volatiles and dust released would provide a greater understanding of the composition of the nucleus itself. The Centaur 29P/S-W1 is the most compelling target for a spacecraft mission because it is renowned for its considerable outburst activity (e.g., Miles et al., 2016) that is remarkably constant and has already provided tantalizing clues to the volatile composition of this Centaur.

The application of advanced observational techniques, including mass spectrometry for isotopic and noble gas analyses, combined with new telescopes like JWST and the Vera Rubin Observatory, promise to revolutionize our understanding of Centaur volatiles. JWST can provide valuable insights into the composition and activity of Centaurs, especially those exhibiting volatile-rich surfaces. Long-term observations of select Centaurs will help us to understand how Centaurs change as they transition between their distant orbits and closer encounters with the Sun. By comparing observations taken at various points in their orbits, it allows us to understand how factors such as solar distance, temperature, and radiation influence their behavior and surface characteristics.

# Acknowledgements

K.E.M. acknowledges support from NASA ROSES RDAP grant 80NSSC19K1306. O.I. was supported by the Slovak Academy of Sciences (grant Vega 2/0059/22) and by the Slovak Research and Development Agency under Contract no. APVV-19-0072. N.X.R. acknowledges support by the Planetary Science Division Internal Scientist Funding Program through the Fundamental Laboratory Research (FLaRe) work package. D.Z.S. is supported by an NSF Astronomy and Astrophysics Postdoctoral Fellowship under award AST-2303553. This research award is partially funded by a generous gift of Charles Simonyi to the NSF Division of Astronomical Sciences. The award is made in recognition of significant contributions to Rubin Observatory's Legacy Survey of Space and Time.